\documentclass[a4paper,11pt]{article}
\pdfoutput=1 

\usepackage{jcappub} 

\usepackage[T1]{fontenc} 
\usepackage{subcaption} 
\usepackage{graphicx}
\usepackage{microtype}
\usepackage{multirow}
\usepackage{mathtools}
\usepackage{epstopdf}
\usepackage[normalem]{ulem}
\usepackage{siunitx}
\DeclareSIUnit \parsec {pc}
\DeclareSIUnit \year {yr}
\usepackage{amsmath}

\def\nn{\nonumber}

\def\Mpl{M_{\rm pl}}
\def\d{{\rm d}}

\title{\boldmath New constraints on axion-gauge field dynamics during inflation from \textit{Planck} and BICEP/Keck data sets}

\author[a,b,1]{Paolo Campeti,\note{Corresponding author.}}
\author[c]{Ogan Özsoy,}
\author[a]{Ippei Obata}
\author[d]{and Maresuke Shiraishi}

\affiliation[a]{Max Planck Institute for Astrophysics, Karl-Schwarzschild-Str.1, 85741 Garching, Germany}
\affiliation[b]{Excellence Cluster ORIGINS, Boltzmannstr. 2, 85748 Garching, Germany}
\affiliation[c]{CEICO, Institute of Physics of the Czech Academy of Sciences, Na Slovance 1999/2, 182 21, Prague.}
\affiliation[d]{Department of General Education, National Institute of Technology,
Kagawa College, 355 Chokushi-cho, Takamatsu, Kagawa 761-8058, Japan}

\emailAdd{pcampeti@mpa-garching.mpg.de}
\emailAdd{ozsoy@fzu.cz}
\emailAdd{obata@mpa-garching.mpg.de}
\emailAdd{shiraishi-m@t.kagawa-nct.ac.jp}

\abstract{We present new constraints on spectator axion-${\rm U}(1)$ gauge field interactions during inflation using the latest \textit{Planck} (PR4) and BICEP/Keck 2018 data releases. This model can source tensor perturbations from amplified gauge field fluctuations, driven by an axion rolling for a few e-folds during inflation. The gravitational waves sourced in this way have a strongly scale-dependent (and chiral) spectrum, with potentially visible contributions to large/intermediate scale $B$-modes of the CMB. We first derive theoretical bounds on the model imposing validity of the perturbative regime and negligible backreaction of the gauge field on the background dynamics.   
Then, we determine bounds from current CMB observations, adopting a frequentist profile likelihood approach. We study the behaviour of constraints for typical choices of the model's parameters, analyzing the impact of different dataset combinations. 
We find that observational bounds are competitive with theoretical ones and together they exclude a significant portion of the model's parameter space. We argue that the parameter space still remains large and interesting for future CMB experiments targeting large/intermediate scales $B$-modes.}

\begin{document}
\maketitle
\flushbottom

\section{Introduction}
\label{sec:intro}
A stochastic gravitational wave background (hereafter SGWB) at all frequencies is a generic prediction of the inflationary paradigm \cite{Grishchuk:1974ny, Starobinsky:1979ty}. The importance of measuring such primordial signal cannot be overstated, since a detection would provide strong evidence for cosmic inflation \cite{Guth:1980zm, Sato:1980yn, Linde:1981mu, Albrecht:1982wi}. To reach this goal, an extensive experimental effort, targeting the SGWB spectrum at different frequencies, is ongoing and will continue throughout the next decade and beyond (see e.g. \cite{Campeti:2020xwn} for a review).   

The amplitude of the SGWB is usually parametrized by the ratio of the amplitudes of the tensor and scalar modes power spectra, the tensor-to-scalar ratio $r$. 
Currently, only upper bounds exist on $r$, the tightest one being $r<0.034$ at $95\%$ C.L.\footnote{The upper limit tightens to $r<0.032$ when adding BAO and CMB lensing data \cite{Tristram:2021tvh}.}
\cite{Tristram:2021tvh} 
placed using a combination of \textit{Planck} \cite{Tristram:2020wbi} and BICEP/Keck \cite{BICEP_2021} CMB experiments data, through the imprint primordial tensor modes are known to leave in the $B$-mode polarization pattern of the CMB \cite{Kamionkowski:1996ks, Seljak:1996gy}.  
Being the most sensitive and the closest in the future timeline among all planned probes of the primordial SGWB \cite{Campeti:2020xwn}, a positive gravitational wave (GW) detection is likely to come first from future CMB $B$-mode experiments, such as the LiteBIRD satellite \cite{Hazumi:2019lys} and the ground-based CMB-S4 \cite{CMB-S4:2016ple}, if tensor-to-scalar ratio reaches $r\sim 10^{-3}$ at CMB scales. 

Merely detecting $r$, however, does not immediately enable us to discriminate between different possible origins of the primordial SGWB. In the simplest scenario of single field inflation, realized by a slowly rolling scalar field minimally coupled to gravity, the SGWB is produced by the quantum vacuum fluctuations of the metric \cite{Grishchuk:1974ny, Starobinsky:1979ty}. 
In this simple setup, the tensor-to-scalar ratio $r$ can be related directly to the energy scale of inflation \cite{Lyth:1996im}.
Furthermore, the SGWB spectrum produced within this  framework is known to be $(i)$ nearly scale-invariant (with a slight red-tilt), $(ii)$ nearly Gaussian and $(iii)$ non-chiral (i.e. parity-conserving). 
However, the relation between the energy scale of inflation and $r$, together with all the properties enunciated above, can be violated if an energetically-excited extra particle content is present during inflation, feeding the stress-energy tensor in the perturbed Einstein equation for the tensor modes of the metric. This intriguing possibility makes testing the scale dependence, Gaussianity and the chirality of the SGWB very compelling, if $r$ is detected by future experiments~\cite{Komatsu:2022nvu}.

Sourcing gravitational waves with additional matter fields during inflation can have, however, undesirable side effects: the sources are always (at least) gravitationally coupled to the sector responsible for the accelerated expansion, and therefore enhance not only tensor but also strongly non-Gaussian scalar modes (see e.g. \cite{Barnaby:2010vf,Mirbabayi:2014jqa,Ferreira:2014zia,Ozsoy:2014sba}). Avoiding overproduction of such perturbations to comply with the tight bounds enforced by the CMB data on non-Gaussianity, while simultaneously maintaining a visible amplitude for the sourced SGWB signal, is therefore a necessary ingredient of any successful inflationary model aiming to achieve GWs of secondary/non-vacuum origin.

Two of the most studied mechanisms that are capable of successfully realizing the above scenario, involve production of a SGWB from amplification of gauge fields of the Abelian \cite{Sorbo:2011rz, Barnaby:2010vf, Barnaby:2012xt, Cook:2013xea, Cook:2011hg, Namba:2015gja, Shiraishi:2016yun, Domcke:2016bkh, Ozsoy:2020ccy, Choi:2015wva, Fujita:2018zbr, Kawasaki:2019hvt,Ozsoy:2020kat} and  non-Abelian \cite{Maleknejad:2011jw, Maleknejad:2011sq, Maleknejad:2012fw, Maleknejad:2016qjz, Dimastrogiovanni:2012ew, Dimastrogiovanni:2016fuu, Obata:2016tmo, Agrawal:2017awz, Agrawal:2018mrg, Adshead:2013qp, Adshead:2013nka, Adshead:2016omu, Adshead:2017hnc} kind.
Particle production of the gauge fields from inflation has been considered in the contexts of several cosmological phenomena, such as the generation mechanisms of magnetic field \cite{Ratra:1991bn,Garretson:1992vt,Anber:2006xt,Martin:2007ue,Demozzi:2009fu,Durrer:2010mq,Caprini:2014mja,Fujita:2015iga,Adshead:2016iae,Fujita:2019pmi,Maity:2021qps} and the matter-antimatter asymmetry in our universe \cite{Alexander:2011hz,Maleknejad:2014wsa,Fujita:2016igl,Caldwell:2017chz,Jimenez:2017cdr,Domcke:2019mnd,Maleknejad:2020yys,Maleknejad:2020pec}, among many others.
Specifically, in this work we focus on the sourcing of GWs through the Abelian ${\rm U}(1)$ gauge field fluctuations and in this context we consider a spectator sector including a generic pseudo Nambu-Goldstone boson (i.e. an axion field) coupled to an Abelian gauge field \cite{Namba:2015gja, Ozsoy:2020ccy}. In this setting, inflation is realized through a standard inflaton sector minimally coupled to gravity and whose energy density is dominant with respect to that of the spectator axion and gauge field, thus allowing controlled production of scalar perturbations. In order to break the conformal invariance in the gauge field sector and allow for the amplification of the fluctuations in the latter with subsequent enhancement of tensor modes during inflation, the axion and the ${\rm U}(1)$ field are considered to interact through a Chern-Simons term \cite{Ni:1977zz, Turner:1987bw}. The resulting gauge field amplification (and also its impact on the scalar curvature perturbations) is controlled by the transient rolling of the axion along its potential. In particular, for the specific realizations of this scenario, we will consider two different choices for the spectator axion potential, both leading to localized gauge field amplification: i.e one with a standard cosine-type potential \cite{Freese:1990rb, Namba:2015gja}; and the other endowed with a string-inspired non-compact\footnote{See e.g. \cite{McAllister:2008hb, McAllister:2014mpa} for an explicit realization of such potentials from the top-down perspective.} axion potential \cite{Ozsoy:2020ccy}. Excitingly, due to parity violating nature of Chern-Simons interaction, only one of the two helicities of the gauge field fluctuations is amplified by the rolling spectator axion, resulting in a fully chiral SGWB. 

Gauge field sources leave distinctive signatures in the primordial SGWB compared to the standard quantum fluctuations of the metric. More specifically, we can exploit the characteristic scale-dependence of sourced tensor modes to distinguish between the two \cite{Namba:2015gja, Barnaby:2011qe, Ozsoy:2020ccy, Cook:2011hg}. Another possibility is to look for the strongly non-Gaussian signature in the bispectra of sourced gravitational waves at CMB \cite{Barnaby:2010vf,Barnaby:2011vw, Shiraishi:2016yun, Shiraishi:2019yux} and interferometer scales \cite{Barnaby:2011qe, Bartolo:2016ami}. Finally, amplification of ${\rm U}(1)$ gauge field sources during inflation is a parity-breaking process and therefore non-zero parity-violating correlations in the CMB angular power spectra \cite{Lue:1998mq, Sorbo:2011rz} and bispectra \cite{Cook:2013xea,Shiraishi:2016yun, Shiraishi:2019yux,Ozsoy:2021onx} are expected, together with circularly polarized gravitational waves at scales relevant for laser interferometers \cite{Thorne:2017jft}. 

In this work, we are going to focus our attention to the scale-dependent signatures of the spectator axion-${\rm U}(1)$ gauge field dynamics during inflation in the angular power spectra at large/intermediate CMB scales\footnote{Indeed, at these scales non-Gaussianity bounds from the CMB are weaker and can be evaded for an axion rolling for just a few e-folds during inflation \cite{Namba:2015gja}. See also the discussion in the beginning of Section \ref{sec:observation}.}, with the aim of deriving constraints on the parameter space of spectator axion gauge field interactions. For this purpose, we utilize the state-of-the art CMB dataset for temperature and polarization provided by the \textit{Planck} satellite \cite{Planck:2020olo} in its latest incarnation, complemented by the latest BICEP/Keck $B$-mode polarization data \cite{BICEP_2021}. In particular, \textit{Planck} yields the current best measurement of temperature and ($E$ and $B$-mode) polarization at the largest scales, which are in fact accessible only from space, while the ground-based BICEP/Keck contributes with the best measurement of intermediate scales $B$-modes to date.       

We explore the likelihoods using a frequentist method: the profile likelihood. Despite being widespread within the particle physics community \cite{ParticleDataGroup:2020ssz, ATLAS:2013mma}, this statistical technique is relatively less common in cosmology\footnote{One of the notable case of use  in cosmology is the application to \textit{Planck} data \cite{planck_profile}, where a comparison was drawn between parameter estimates from profile likelihoods and the usual Bayesian Monte Carlo Markov Chain (MCMC) method.}, compared to Bayesian methods. Nonetheless, profile likelihoods present several advantages over the latter: the global maximum likelihood solution is guaranteed by construction and, moreover, parameters estimates are independent from prior distributions and model parametrization, thus making profile likelihoods immune to ``volume effects'', appearing during marginalization \cite{Hamann:2007pi}.     

The structure of the paper is as follows. We start in Section \ref{sec:theory} by reviewing the spectator axion-${\rm U}(1)$ model and assessing the bounds on its parameter space imposed by theoretical self-consistency (namely from perturbativity and backreaction considerations, see subsection \ref{s1p2}).   
In Section \ref{sec:observation} we provide instead the latest observational bounds on the model parameters from \textit{Planck} and BICEP/Keck data. We offer a detailed interpretation of the constraints for different choices of the model parameters, highlighting differences among the two axion potentials, and analyze the impact of different dataset combinations.
Finally, in Section \ref{sec:conclusions}, we show that observational constraints on the model are competitive with theoretical ones and update the available parameter space of the model. We conclude by suggesting a possible future path for discriminating the spectator axion-${\rm U}(1)$ model we consider from the conventional single field realizations of inflation at CMB scales. 

\section{Theory: A spectator axion-U(1) gauge field model}\label{sec:theory}

As we mentioned in the introduction, the particle content we consider during inflation contains a spectator axion-Abelian $\mathrm{U}(1)$ gauge field sector along with a canonical inflaton sector that does not exhibit direct interactions with the former and both sectors minimally coupled to gravity. The action describing this system is given by \cite{Barnaby:2012xt,Mukohyama:2014gba,Namba:2015gja},
\begin{equation}\label{AfM}
S = \int \d^4 x { \sqrt{-g}} \,\bigg[ \frac{\Mpl^2R}{2} -\underbrace{\frac{1}{2}(\partial\phi)^2-V(\phi)}_{\equiv\, \mathcal{L}_{\rm inflaton}} -\underbrace{ \dfrac{1}{2}(\partial\chi)^2 - U(\chi) -\dfrac{1}{4}F_{\mu\nu}F^{\mu\nu}+ \mathcal{L}_{\rm int} }_{\equiv \, \mathcal{L}_{\rm spectator}}\bigg],
\end{equation}

\noindent where the first term represents the standard Einstein-Hilbert action, $\phi$ is the inflaton and the spectator sector includes an axion-like field $\chi$, ${\rm U}(1)$ gauge field $A_\mu$ with the antisymmetric field strength tensor $F_{\mu\nu} =\partial_{\mu} A_{\nu} - \partial_{\nu} A_{\mu}$.  In the spectator sector $\mathcal{L}_{\rm spectator}$, we consider an axion-like field $\chi$ that enjoys a(n) (approximate) shift symmetry, taking into account  the gauge invariance of the ${\rm U}(1)$ field, $\chi$ is then expected to interact with gauge fields through the leading order dimension five operator\footnote{Another possibility is the coupling of a shift symmetric scalar to the fermion current via dimension five operator, see e.g. \cite{Adshead:2015kza,Adshead:2018oaa,Adshead:2019aac} for phenomenological implications of this scenario.} that takes the following form

\begin{equation}\label{Lint}
\mathcal{L}_{\rm int} = -\lambda\dfrac{\chi}{f}F_{\mu\nu}\tilde{F}^{\mu\nu},
\end{equation}
where $\tilde{F}^{\mu\nu} \equiv \sqrt{-g}\,\epsilon^{\mu\nu\rho\sigma}F_{\rho\sigma}/ 2$ is the Hodge dual of the field strength tensor $F_{\mu\nu}$, $f$ is the axion decay constant, $\lambda$ is a dimensionless coupling constant and $\epsilon^{\mu\nu\rho\sigma}$ is an anti-symmetric tensor satisfying $\epsilon^{0123} = g^{-1}$.

\noindent{\bf Background evolution.}
We consider an inflationary setup where the spectator sector fields provide subleading contribution to the total energy density during inflation. This implies that energy densities of the scalar fields in the model \eqref{AfM} obey $\rho_\phi \gg \rho_\chi$ where $\rho_X = \dot{X}^2/2 + V(X)$ (with an overdot denoting derivative with respect to cosmological time $t$) for $X = \{\phi, \chi\}$ and assuming negligible backreaction from gauge field fluctuations $\rho_A \ll \rho_\chi$ (see Section \ref{s1p2}), we have 

\begin{equation}
3H^2\Mpl^2 \simeq \rho_\phi + \rho_\chi \quad \longrightarrow \quad 3H^2\Mpl^2 \simeq V(\phi),
\end{equation}
such that quasi-dS expansion is completely dictated by the inflaton's potential. Furthermore, we assume that $V(\phi)$ is flat enough to support sufficiently long quasi dS expansion but otherwise we let it unspecified as the fine details regarding the inflaton's dynamics is irrelevant for the discussion we present below. With this assumption, we will treat Hubble rate $H$ as constant and denote the scale factor during inflation in conformal time as 
$a(\tau) = -1/ (H\tau)$\,\,\footnote{In this work, we disregard terms that are subleading in slow-roll expansion.}, $-\infty<\tau\leq 0$. 
On the other hand, if the spectator axion $\chi$ is displaced from its global minimum, it can also roll down in its potential, albeit in the slow-roll regime thanks to sufficiently flat potential $U(\chi)$ endowed with shift symmetry. Due to the slow-roll assumption, we require that $\chi$'s background dynamics should obey
\begin{equation}\label{src}
\bigg|\frac{\ddot{\chi}}{3H\dot{\chi}}\bigg| \ll 1,
\end{equation}
during inflation when scales associated with CMB observations exits the horizon. In what follows, we will briefly review the impact of such a slowly rolling spectator axion on the behavior of gauge field fluctuations. For the clarity of the discussion, initially we will not specify the explicit form of the potential $U(\chi)$ before we introduce explicit axion models (see Section \ref{rams}) that we analyze in this work.  

\noindent{\bf Amplification of gauge field fluctuations.} For a dynamical spectator axion field with a time dependent profile $\chi(t)$, the interaction term \eqref{Lint} $\mathcal{L}_{\rm int} \propto \chi(t) \partial_{\mu}(A_\nu (\partial_\rho A_\sigma))$ can no longer be treated as a surface term in the action \eqref{AfM}. As a result, the dispersion relation of $A_\mu$ becomes modified, leading to copious production of its fluctuations provided that spectator axion has a non-trivial velocity $\dot{\chi}\neq 0$. To see this, we first decompose gauge field fluctuations into Fourier modes in Coulomb gauge ($A_0 = 0$) as \cite{Anber:2009ua},
\begin{equation}\label{Adecom}
    \hat{A}_{i}(\tau, \vec{x})=\sum_{\lambda=\pm} \int \frac{\mathrm{d}^{3} k}{(2 \pi)^{3 / 2}}\left[\epsilon_{i}^{(\lambda)}(\vec{k}) A_{\lambda}(\tau,k) \hat{a}_{\lambda}(\vec{k}) \mathrm{e}^{i \vec{k} \cdot \vec{x}}+\mathrm{h}. \mathrm{c}\right],
\end{equation}
where h.c represent hermitian conjugate of the first term in \eqref{Adecom} and the helicity vectors obey $k_i \epsilon^{\pm}_i = 0$, $\epsilon_{ijk}~ k_j ~\epsilon^{\pm}_k = \mp i k \epsilon^{\pm}_i$, $\epsilon^{\pm}_i\epsilon^{\pm}_i = 0$, $\epsilon^{\pm}_i \epsilon^{\mp}_i =1$, $(\epsilon^{\lambda}_i(\vec{k}))^{*} = \epsilon^{\lambda}_i(-\vec{k}) = \epsilon^{-\lambda}_i(\vec{k})$ together with the commutation relations of annihilation/creation operators $\left[\hat{a}_\lambda(\vec{k}),\hat{a}^\dagger_{\lambda'}(\vec{k}')\right] = \delta_{\lambda\lambda'} \,\, \delta(\vec{k}-\vec{k}')$. 

Inserting the decomposition \eqref{Adecom} in the spectator part of the action \eqref{AfM}, the equation of motion (EoM) for the gauge field mode functions in a flat FLRW background satisfy 
\begin{equation}\label{mea}
\partial_x^2 A_\pm + \left(1 \pm \dfrac{2\xi}{x}\right)A_\pm = 0 \ , \quad \xi \equiv -\dfrac{\lambda\dot{\chi}}{2Hf} \ , 
\end{equation}
where we defined dimensionless time variable $x \equiv -k\tau$ and the effective coupling $\xi$ between spectator axion and gauge field. Without any loss of generality, we work with $\xi > 0$ and $\dot{\chi}< 0$ so that spectator axion rolls down on its potential from positive large to small values $\chi \geq 0$. As we mentioned before, the correction that appear in the dispersion relation \eqref{mea} arise through the coupling \eqref{Lint} whose parity violating nature is apparent from its alternating sign $\pm$. In particular, when the modes are deep inside the horizon ($x = k/(aH) \gg 1$), the correction term is negligible and the gauge field obeys the standard dispersion relation. However, as the modes stretches outside the horizon, it becomes dominant for $x = k/(aH) \lesssim 2\xi$, leading to instability for one of the circular polarization state of the gauge fields. In our conventions ($\xi > 0$ \& $\dot{\chi} < 0$), $A_-$ state experiences a tachyonic instability while $A_+$ stays in its vacuum. The roll of spectator axion field $\dot{\chi} \neq 0$ therefore induces production of helical gauge fields and assuming the roll of axion with a constant velocity $\xi \simeq constant$, the late time ($-k\tau \to 0$) gauge field mode functions amplified this way typically exhibit an exponential amplitude that is regulated by the axion's velocity  \cite{Anber:2009ua}:
\begin{equation}\label{GFA}
A_- \propto e^{\pi \xi} = e^{\pi {\lambda |\dot{\chi}|}/{(2Hf)}} .
\end{equation} 

In the model \eqref{AfM}, although the spectator sector  does not exhibit direct couplings with the visible sector fluctuations such as the inflaton $\delta\phi$ and metric  $h_{ij}$, the influence of the particle production processes in the gauge fields inevitably mediate to the visible sector perturbations through gravitational interactions. Below, we will briefly review the impact of the gauge field sources on these fluctuations. 

\noindent{\bf Tensor perturbations sourced by vector fields.} To study the influence of gauge field fluctuations on the tensor perturbations, we focus on the transverse traceless metric perturbation  $g_{ij} = a^2(\tau) [\delta_{ij} + \hat{h}_{ij}(\tau,\vec{x})]$ and decompose it into circularly polarized states $\lambda = \pm$ in Fourier space as $\hat{h}_\lambda (\tau, \vec{k}) = \Pi_{ij,\lambda} (\vec{k})\,\hat{h}_{ij}(\tau,\vec{k})$ where $\Pi_{ij,\lambda}$ is the polarization tensor obeying $\hat{k}_i\, \Pi_{ij,\lambda}(\vec{k})=0 $,  $\Pi^{*}_{ij,\lambda}\Pi_{ij,\lambda'} = \delta_{\lambda\lambda'}$ and $\Pi^{*}_{ij,\lambda}(\vec{k}) = \Pi_{ij,-\lambda}(\vec{k}) = \Pi_{ij,\lambda}(-\vec{k})$. Expanding the action \eqref{AfM} up to third order in fluctuations including $\hat{A}_i$ and $\hat{h}_{ij}$, it can be shown that the mode equation of graviton polarization states $h_{\lambda}$ is sourced by the transverse, traceless part of the energy momentum tensor that is composed of gauge field fluctuations \cite{Barnaby:2012xt}:
\begin{equation}\label{te}
\left(\partial^2_\tau + k^2 -\frac{2}{\tau^2}\right)(a \hat{h}_\lambda) =-\frac{2a^{3}}{\Mpl^2} \Pi_{i j, \lambda}(\vec{k}) \int \frac{\d^{3} x}{(2 \pi)^{3 / 2}} \mathrm{e}^{-i \vec{k} \cdot \vec{x}}\left[\hat{E}_{i} \hat{E}_{j}+\hat{B}_{i} \hat{B}_{j}\right]
\end{equation}
where we defined dark ``electric" and ``magnetic" fields $\hat{E}_i (\tau, \vec{x}) = -{a^{-2}} \hat{A}_{i}^{\prime}(\tau, \vec{x}), \,\, \hat{B}_{i}={a^{-2}} \epsilon_{i j k} \partial_{j} \hat{A}_{k}$ in a flat FLRW universe, in analogy with standard model electromagnetism.

\smallskip
\noindent{\bf Scalar perturbations sourced by vector fields.} The impact of particle production processes on the visible scalar fluctuations is also encoded indirectly by the presence of gravitational interactions \cite{Ferreira:2014zia}. In particular, integrating out the non-dynamical scalar metric fluctuations such lapse $\delta N$ and the shift $N^{i}$ reveals a mass mixing between inflaton $\delta \phi$ and spectator axion $\delta \chi$ fluctuations and opens up a channel that can influence the curvature perturbation
\footnote{Regarding other direct contributions to $\mathcal{R}$ from fluctuations in the spectator axion $\delta \chi$ and gauge fields $A_i$, in this work we will consider a spectator axion model that rolls down to its minimum long before the end of inflation (see Section \ref{rams}), so that the contribution of $\delta \chi$ on the late time curvature perturbation $\mathcal{R}$ can be neglected \cite{Mukohyama:2014gba,Namba:2015gja}. On the other hand, the contribution from gauge fields is proportional to the absolute value of Poynting vector, $a|\vec{S}| = a|\vec{E}\times \vec{B}|$ which is also negligible at late times as the particle production saturates at super-horizon scales and the resulting electromagnetic fields decay as $\vec{E},\vec{B} \sim a^{-2}$ \cite{Ozsoy:2020ccy}. For the purpose of evaluating phenomenological implications of the model \eqref{AfM}, we therefore adopt the standard relation $\mathcal{R} \equiv  H \delta \phi/\dot{\phi}$ in this work.} $\mathcal{R} \simeq {H\,\delta \phi}/{\dot{\phi}} $ through the inverse decay of gauge fields: $A_i + A_i \to \delta \chi \to \delta \phi \propto \mathcal{R}$. Dynamics of this contribution can be understood by first studying the influence of particle production on the spectator axion fluctuations $\delta \chi$ through,
\begin{equation}
\label{usgm}\left(\frac{\partial^{2}}{\partial \tau^{2}}+k^{2}-\frac{2}{\tau^{2}}\right) (a \delta \hat{\chi}) \simeq \frac{a^{3}\lambda}{f} \int \frac{\d^{3} x}{(2 \pi)^{3 / 2}} \mathrm{e}^{-i \vec{k} \cdot \vec{x}} \, \hat{E}_i(\tau, \vec{x}) \hat{B}_i(\tau, \vec{x}).
\end{equation}
Then focusing on the inhomogeneous solution of the $\delta \chi$ fluctuations in \eqref{usgm}, one can compute the conversion of $\delta \chi$ to $\delta \phi$ via 
\begin{equation}
\label{uphi} \left(\frac{\partial^{2}}{\partial \tau^{2}}+k^{2}-\frac{2}{\tau^{2}}\right) (a \delta \hat{\phi}) \simeq 3a^2 \frac{\dot{\phi}\dot{\chi}}{\Mpl^2} ~(a \delta \hat{\chi}),
\end{equation}
to find the the part of curvature perturbation $\mathcal{R}$ that is sourced by the amplified gauge fields.

In this work, we are interested to the extent of which gauge field sources can influence tensor perturbations consistent with other observations at CMB scales. In this context, it has been recently realized that if the spectator axion $\chi$ rolls for a large-amount of e-folds $\Delta N_\chi \gg 1$ during which the scales associated with CMB observations exit the horizon\footnote{The regime of validity of the perturbative description of gauge field production is also questioned in \cite{Ferreira:2015omg,Peloso:2016gqs}.}, the sourced contributions to the $\mathcal{R}$ becomes sizeable due to the sensitivity of gauge field amplitudes and $\delta\phi-\delta\chi$ mass mixing (see eq. \eqref{uphi}) on the spectator axion's velocity $\xi \propto |\dot{\chi}|$ \cite{Ferreira:2014zia}. In particular, insisting on a large secondary contribution to the tensor fluctuations through eq. \eqref{te} generically  leads to an exceedingly large scalar non-Gaussianity at CMB scales \cite{Ferreira:2014zia,Ozsoy:2017blg}. 
The origin of the difficulty in efficiently enhancing tensor perturbations compared to the scalars can be readily seen from eqs. \eqref{te}, \eqref{usgm} and \eqref{uphi}, by realizing that the sourced part of both perturbations arise via non-linear terms including the same amount of gauge fields. Notice however from eqs. \eqref{usgm} and \eqref{uphi} that the efficiency of the process  $A_i + A_i \to \delta \chi \to \delta \phi \propto \mathcal{R}$ is highly sensitive to the behavior of the spectator axion's velocity as both gauge field production (see eq. \eqref{GFA}) and mass mixing $\delta\phi-\delta\chi$ have dependence on $\dot{\chi}$. In what follows, we will discuss two spectator axion models that exhibit a localized velocity profile that can overcome the aforementioned limitations on scalar fluctuations.  

\subsection{Transiently rolling spectator axion models}\label{rams}
In order to minimize the influence of particle production on the curvature perturbation and to render secondary GWs sourced by gauge fields viable, we will consider models that can lead to localized gauge field production where the spectator axion transiently rolls on potentials of the following form \cite{Namba:2015gja,Ozsoy:2020ccy}: 
\begin{equation}\label{pots}
U(\chi)=
 \begin{dcases} 
       \Lambda^4 \left[1-\cos\left(\frac{\chi}{f}\right)\right],& \quad{\rm Model\, 1}\,({\rm M}1) \,,\\
        \mu^3\chi + \Lambda^4 \left[1-\cos\left(\frac{\chi}{f}\right)\right]\,\,\&\,\, \Lambda^4\lesssim \mu^3 f &\quad{\rm Model\, 2}\,({\rm M}2),
   \end{dcases}
\end{equation}
where $\mu$ and $\Lambda$ are parameters of mass dimension one. 

The first model (M1) features a standard shift symmetric potential (see e.g. \cite{Freese:1990rb}) with the size of the axion modulations is set by $\Lambda$. In this model, the motion of the axion is contained within the maximum ($\chi = \pi f$) and the minimum ($\chi = 0$) where the slope $U'(\chi)$ vanishes. Therefore at large (early times) and small field values (late times), axion rolls with very small velocities whereas $\dot{\chi}$ obtains relatively large value at an intermediate time when $\chi$ passes through an inflection point $\chi_* = \chi(\tau_*)$ with $U''(\chi_*) = 0$ where the slope of the potential $U'(\chi)$ becomes maximal. 
\begin{figure}[t!]
\begin{center}
\includegraphics[scale=0.825]{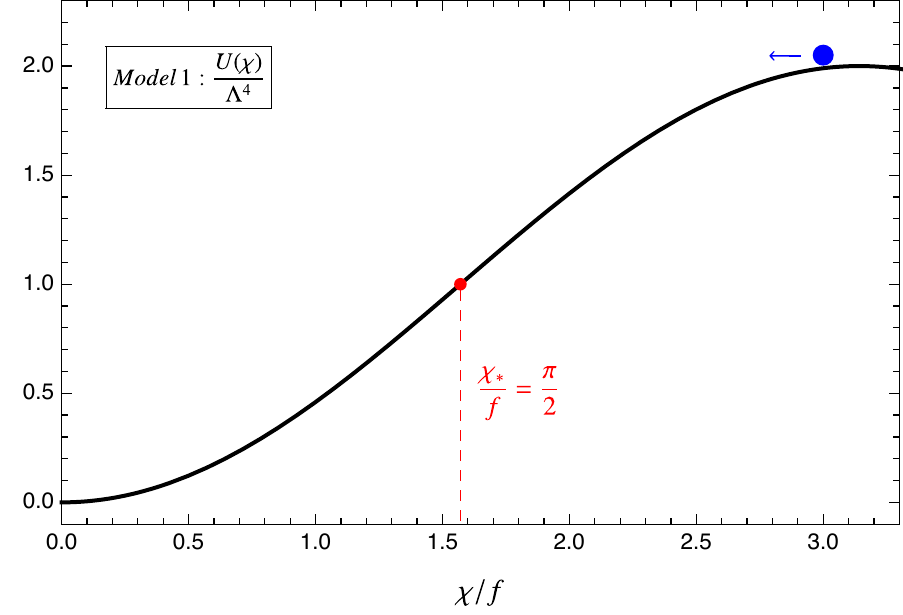}\includegraphics[scale=0.81]{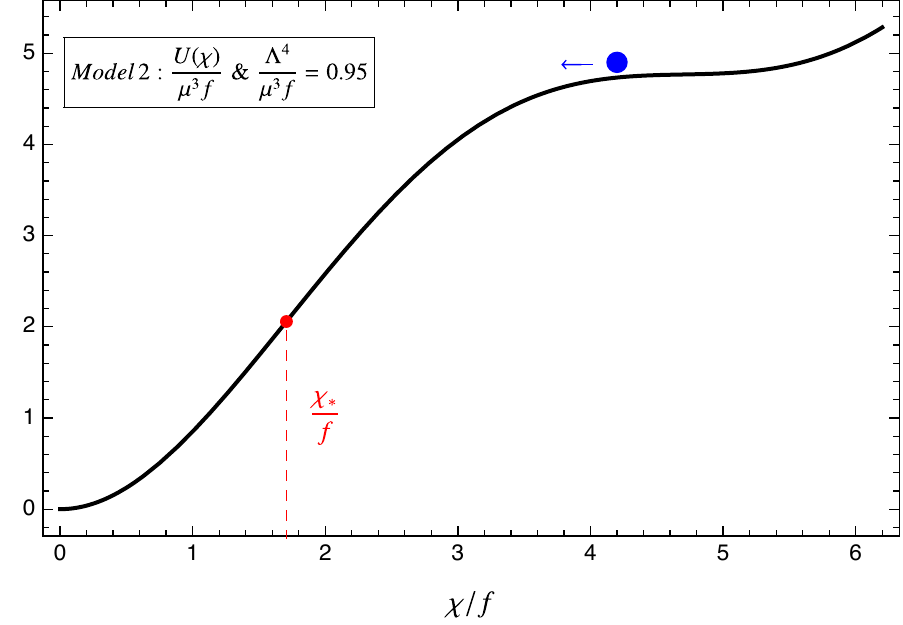}
\end{center}
\caption{The profile of spectator axion potentials \eqref{pots} for M1 (Left) and M2 (Right). For both panels red dot locates the position of the inflection point where the slope of the potential $U'(\chi)$ becomes maximal. 
\label{fig:ram}}
\end{figure}

In the second model (M2), the axion field range is extended via a monodromy term \cite{McAllister:2008hb,McAllister:2014mpa} proportional to a soft symmetry breaking mass parameter $\mu$ and $\chi$ is assumed to probe step-like feature(s)\footnote{In the bumpy regime, depending on the initial conditions ($\chi \gg f$) spectator axion can probe multiple step-like features during inflation. In this work, we assume that $\chi$ traverse only one such region on its potential during which observable scales associated with CMB exits the horizon.} in the ``bumpy" regime, $\Lambda^{4}\lesssim \mu^3 f$. Similarly to the first model, in the plateau like region and towards the global minimum\footnote{As shown in the right panel of Fig. \ref{fig:ram}, the roll of $\chi$ towards the global minimum ($\chi = 0$) can be captured by modifying the monomial term as $\mu^3 \chi \to \mu^3 f [\sqrt{1 + (\chi/f)^2} -1 ]$ so that the axion potential \eqref{pots} interpolates between $\mu^3 \chi$ and $(\mu^3/f) \chi^2$ from large to small field ($\chi/f \to 0$) values respectively. By construction this modification is designed to affect the motion of $\chi$ far away from the inflection point $\chi_*$ where the axion's velocity peaks and therefore we do not expect it to influence $\chi$'s velocity profile \eqref{Joep} and the resulting phenomenology we discuss in this work (see below) which are derived by assuming the potential shown in \eqref{pots}. In short, it only stands to ensure a smooth rollover to the minimum $\chi \to 0$ with a small velocity $U' \propto \dot{\chi} \to 0$ within the slow-roll approximation we are undertaking.}, the spectator axion acquires very small velocities where $U' \to 0$ but obtains a transient peak when the slope of the potential $U'(\chi)$ becomes maximal at the cliff region, in particular at inflection point denoted by $\chi_*$ (See Fig. \ref{fig:ram}).    

Assuming slow-roll motion \eqref{src}, for typical field ranges dictated by the scalar potentials in \eqref{pots}, the spectator field velocity $\dot{\chi}$ and the effective coupling strength $\xi = -\lambda\dot{\chi}/(2Hf)$ therefore obtains a peaked time dependent profile \cite{Namba:2015gja,Ozsoy:2020ccy}:
\begin{equation}\label{Joep}
\xi(\tau)=
\begin{dcases} 
       \,\frac{2 \xi_{*}}{\left({\tau_*}/{\tau}\right)^{\delta}+\left({\tau}/{\tau_*}\right)^{\delta}},& \,\,  \,\,\, \delta \equiv \frac{\Lambda^4}{3H^2 f^2}\quad({\rm M}1) \\
     \, \frac{\xi_{*}}{1+\ln[\left({\tau}/{\tau_*}\right)^{\delta}]^2},& \,\, \,\,\, \delta \equiv \frac{\mu^3}{3H^2 f}\quad\,\,({\rm M}2)
\end{dcases}
\end{equation}

where the subscript $*$ denotes the value of a quantity at the time when the axion passes through the inflection point (See Fig. \ref{fig:ram}). It is clear from \eqref{Joep} that $\xi$ obtains its peak value at $\tau = \tau_*$ with a maximal value that can be parametrized in terms of the dimensionless coupling constant $\lambda$ as
\begin{equation}\label{Joes}
\xi(\tau_*)\equiv\xi_* =
\begin{dcases} 
      \,\,\, \frac{\lambda \delta}{2},& \,\,\,  \quad({\rm M}1) \\
      \,\,\, \lambda \delta,& \,\, \quad\,\,({\rm M}2).
\end{dcases}
\end{equation}

The width of the time dependent peak in $\xi$ is mainly controlled by the dimensionless ratio $\delta$ that essentially characterize the mass of the spectator axion in its global minimum $\delta \approx m^2_{\chi} /H^2$. In particular, for larger $\delta$ (heavier axion), restoring force towards the global minimum is larger so that $\chi$ traverse the inflection point faster, resulting with a sharper peak in $\xi$. In other words, $\delta$ is a measure for the acceleration ($\dot{\xi}/{(\xi H)} = \ddot{\chi}/(\dot{\chi}H) \sim \delta $) of the spectator axion as it rolls down on its potential. At this point it is worth mentioning that due to the slow-roll approximation \eqref{src} we are undertaking,  we are restricted to $\delta \ll 3$. In this work, to derive observational and theoretical constraints on the spectator axion-gauge field model \eqref{AfM}, we will focus on the following cases $\delta = \{0.2, 0.3, 0.4, 0.6\}$ corresponding to increasingly sharp rise in the velocity of $\chi$. 

The peaked structure of $\xi$ profile introduces a critical scale $\tau_*^{-1} = k_*$ in the EoM \eqref{mea}, corresponding to the scale that exits the horizon when the axion's velocity is maximal (i.e when $\xi = \xi_*$). Since the tachyonic mass of the ${\rm U}(1)$ field in \eqref{mea} is maximal around this point, it results in a scale dependent growth of the gauge field fluctuations where only modes whose size is comparable to the horizon size at $\tau =\tau_*$, i.e $k \sim \mathcal{O}(1) a_* H_*$, are efficiently amplified. The scale dependent amplification of gauge field modes can be accurately studied using the semi-analytic techniques discussed in \cite{Namba:2015gja,Ozsoy:2020ccy}, which is what we will utilize in our analysis. Below we review the impact of such
scale dependent vector field production on the 2-point correlators of tensor and scalar fluctuations
during inflation. 

\noindent{\bf Scale dependent perturbations from gauge field sources.}
In addition to the standard vacuum fluctuations driven by the quasi-dS background, the perturbations in the observable sector $\hat{\mathcal{X}} = \{\hat{\mathcal{R}},\hat{h}_\pm\}$ pick up a sourced contribution from the enhanced gauge field fluctuations that can be described by the particular solutions of \eqref{te} and \eqref{uphi} (see also \eqref{usgm})$:\hat{\mathcal{X}} = \hat{\mathcal{X}}^{(\rm v)} + \hat{\mathcal{X}}^{(\rm s)}$ where the superscripts denote vacuum and sourced modes, respectively. We define the power spectra of $\mathcal{R}$ and $h_\lambda$ as 
\begin{align}
\frac{k^{3}}{2 \pi^{2}} \langle\mathcal{R}(\vec{k}) \mathcal{R}(\vec{k}')\rangle &\equiv \mathcal{P}_{\mathcal{R}}\left(k\right) \,\delta^{(3)}(\vec{k}+\vec{k}') \\
\frac{k^{3}}{2 \pi^{2}} \langle h_{\lambda}(\vec{k}) h_{\lambda'} (\vec{k})\rangle & \equiv \mathcal{P}_{\lambda}({k})\, \delta_{\lambda \lambda'}\, \delta^{(3)}(\vec{k}+\vec{k}'). 
\end{align}

Since the origin of the vacuum and sourced part of scalar and tensor perturbations are different, these contributions are statistically uncorrelated and therefore the resulting total power spectra can be simply described by the sum of vacuum and sourced part auto-correlators as 
\begin{equation}\label{tps}
\mathcal{P}_{\mathcal{R}}(k) = \mathcal{P}^{(\rm v)}_{\mathcal{R}}(k) + \mathcal{P}^{(\rm s)}_{\mathcal{R}}(k),\quad\quad \mathcal{P}_{h}(k) = \sum_{\lambda = \pm}  \left[ \mathcal{P}^{(\rm v)}_{\lambda}(k) + \mathcal{P}^{(\rm s)}_{\lambda}(k) \right] \simeq \mathcal{P}^{(\rm v)}_h(k) + \mathcal{P}^{(\rm s)}_{-}(k),
\end{equation}
where the vacuum contributions are given by the standard expressions:
\begin{equation}\label{psv}
\mathcal{P}_{\mathcal{R}}^{(\rm v)} = \frac{H^2}{8\pi^2\epsilon_\phi \Mpl^2},\quad\quad r_{\rm v} \equiv \frac{\mathcal{P}^{(\rm v)}_{h}}{\mathcal{P}_{\mathcal{R}}^{(\rm v)}} = 16 \epsilon_\phi ,
\end{equation}
with $r_{\rm v}$ denoting the vacuum tensor-to-scalar ratio and $\epsilon_\phi \equiv \dot{\phi}^2/ (2 H^2 \Mpl^2)$ is the slow-roll parameter controlled by the inflaton sector. Note that due to the parity violation in the gauge field sector ($A_{-} \gg A_+$), it is sufficient to take into account $\mathcal{P}^{(\rm s)}_{-}(k)$ in \eqref{tps} among the sourced contributions to tensor modes while the vacuum fluctuations treat both polarization states of the metric democratically: $\mathcal{P}^{(\rm v)}_{+}(k) =\mathcal{P}^{(\rm v)}_{-}(k) $. On the other hand, for the transiently rolling spectator axion models we described above, the sourced power spectra in \eqref{tps} inherit the scale dependence of the gauge field that leads to a Gaussian spectral shape \cite{Namba:2015gja,Ozsoy:2020ccy},
\begin{align}\label{SC}
\nn \mathcal{P}^{(\rm s)}_{j}(k) &=\left[\epsilon_{\phi} \mathcal{P}_{\mathcal{R}}^{(\rm v)}(k)\right]^{2} f_{2, j}\left(\xi_{*},\frac{k}{k_{*}}, \delta\right), \\
f_{2, j}\left( \xi_{*},\frac{k}{k_{*}}, \delta\right) & \simeq f_{2, j}^{c}\left[\xi_{*}, \delta\right] \exp \left[-\frac{1}{2 \sigma_{2, j}^{2}\left[\xi_{*}, \delta\right]} \ln ^{2}\left(\frac{k}{k_* \,x_{2, j}^{c}\left[\xi_{*}, \delta\right]}\right)\right],
\end{align}
where $j = \{\mathcal{R}, \pm\}$. The functions $ f_{2, j}^{c}, \sigma_{2, j}, x_{2, j}^{c}$ control, respectively, the amplitude, the width, and the position of the peak of the sourced signal, which depend on the background model of the spectator axion through the parameters $\xi_*$ and $\delta$ we discussed above and therefore on the underlying scalar potential \eqref{pots} in the spectator axion sector. For representative choices of the background parameter $\delta$, we present accurate formulas for the amplitude $ f_{2, j}^{c}$, width $\sigma_{2, j}$ and the location $x_{2, j}^{c}$ of the peak in terms of the effective coupling $\xi_*$ in Tables \ref{tab:fit1}-\ref{tab:fit4} in the appendix. Notice that due to parity violating nature of gauge field production, sourced tensor perturbations satisfy $f_{2,-}\gg f_{2.+}$ and therefore it is maximally chiral. Note also from the Tables \ref{tab:fit1}-\ref{tab:fit4} that the amplitude $f^c_{2,j}$ of the sourced signals is exponentially sensitive to the effective spectator axion-gauge field coupling $\xi_*$ that parametrizes the efficiency of particle production in the gauge field sector. 

In \eqref{tps}, the sourced power spectra \eqref{SC} introduce a Gaussian bump feature on top of the standard quasi scale-invariant spectra \eqref{psv}. If the former feature is dominant, the total power spectra becomes highly scale-dependent. Clearly, such a scale dependence should not overwhelm the scalar power spectrum and should be consistent with the CMB temperature (T) and polarization modes (E,B) data. Our main goal in this work is to derive constraints on the rolling spectator axion-${\rm U}(1)$ gauge field models from \textit{Planck} and BICEP/Keck 2018 data in order to see to what extent we can realize a chiral, synthetic component of tensor modes through the spectator axion-gauge field dynamics. 

However, before we continue our discussion in this direction, we would like to identify and check the parameter space of the spectator axion-gauge field model \eqref{AfM} consistent with backreaction and perturbativity considerations first discussed in \cite{Ferreira:2015omg,Peloso:2016gqs}.

\subsection{Limits on backreaction and perturbativity}\label{s1p2}
Induced by the gauge field amplification in the spectator sector, the derivation of the scale dependent contributions \eqref{SC} to the total power spectra  assumes that backreaction of the spectator fields on the background evolution is negligible and vector/scalar fluctuations in the spectator sector stay in the perturbative regime. In this section, we study the limitations on the size of the sourced signals from these effects. In our analysis, we will closely follow \cite{Ferreira:2015omg,Peloso:2016gqs,Ozsoy:2020ccy} which we refer the reader for many details presented below. 

\subsubsection{Backreaction constraints}
Since we assume that axion like field is a spectator and does not contribute effectively to the total energy density during inflation, we need to make sure that $\rho_\chi \ll V_{\phi} \simeq 3 H^2 \Mpl^2$ is satisfied during the motion of $\chi$. As shown explicitly in \cite{Peloso:2016gqs,Ozsoy:2020ccy}, for both of the transiently rolling spectator axion models we consider, potential energy of the axion $U({\chi})$ always dominates over the kinetic energy $E_{\rm kin,\chi} = \dot{\chi}^2 /2$ which reaches its maximal value at the inflection point when $\tau = \tau_*$. Therefore, to ensure that spectator $\chi$ does not contribute to the background energy density, it is sufficient to enforce 
\begin{equation}\label{br1}
U(\chi)\big|_{\rm max} \ll 3 H^2 \Mpl^2, 
\end{equation}
where the ``$\rm max$'' refers to the maximum value of the potential energy during the rollover of $\chi$. 
\smallskip

\noindent{\bf An upper bound on $f/\Mpl$:} In the compact spectator axion model, assuming $\chi$ starts its motion close to the maximum $\chi_{\rm in} \simeq \pi/f$ of the potential, the maximal value of the potential energy density is set by the height of the oscillatory potential in  \eqref{pots} and is given by $U(\chi)|_{\rm max} \simeq 2\Lambda^4$. On the other hand, in the non-compact axion model, the maximal value of the potential depends on the initial conditions as $U(\chi)|_{\rm max} \simeq \mu^3 f [ \chi_{\rm in}/f + 1]$ in the $\Lambda^4 \lesssim \mu^3 f$ regime where $\chi_{\rm in} / f \simeq 3\pi/2$, assuming $\chi$ traverse a single cliff-like region in its potential \cite{Ozsoy:2020ccy} (See Fig. \ref{fig:ram}). In terms of the dimensionless ratios we defined in \eqref{pots}, the first condition \eqref{br1} gives
\begin{equation}\label{br1f}
U_{\rm max} \simeq
\begin{dcases} 
  3 H^2 f^2\times \,2\delta & \\
  3 H^2 f^2\times  5.7\delta &
\end{dcases}
\xRightarrow{\text{eq. \eqref{br1}}}\quad
\begin{dcases} 
  \frac{f}{\Mpl}\,\, < \,\,\frac{1}{\sqrt{2\delta}}& \quad({\rm M}1) \\
  \frac{f}{\Mpl}\,\, <\,\, \frac{1}{\sqrt{5.7\delta}}& \quad({\rm M}2).
\end{dcases}
\end{equation}
Therefore, the backreaction constraints translates into an upper bound on the ratio between two fundamental parameters in our model, namely $f/\Mpl$.

\smallskip
\noindent{\bf A lower bound on $f/\Mpl$:} Next, we need to make sure that gauge field production in the spectator sector does not influence the background evolution of $\chi$. Since the gauge field mode functions are amplified at the expense of spectator axion kinetic energy, we therefore require that maximum energy density contained in the gauge fields to be smaller than peak kinetic energy of spectator axion: $\rho_{A, \rm max} \ll (\dot{\chi}^2/2)_{\tau = \tau_*} \simeq \rho_\phi \epsilon_{\chi,*}/3$ where we defined the slow-roll parameter $\epsilon_\chi = \dot{\chi}^2 /(2H^2 \Mpl^2)$. Using the definition of effective coupling in \eqref{mea} and \eqref{Joep}, this condition can be cast into the following form
\begin{equation}\label{br2}
\frac{\rho_{A, \rm max}}{\rho_\phi} \ll \frac{\epsilon_{\chi,*}}{3} \simeq\,\, \begin{dcases} 
  \frac{\delta^2}{6} \left(\frac{f}{\Mpl}\right)^2& \quad({\rm M}1) \\
  \frac{2\delta^2}{3} \left(\frac{f}{\Mpl}\right)^2& \quad({\rm M}2),
\end{dcases}
\end{equation}
where $\rho_{A,\rm max}$ is the maximum value of the gauge field energy density which is typically obtained when $\tau/\tau_* \sim 10^{-2}$ \cite{Peloso:2016gqs,Ozsoy:2020ccy}. For both spectator models we consider, we computed the quantity $\rho_{A, \rm max}/(\rho_\phi \epsilon_\phi)$  using the formulas provided in \cite{Peloso:2016gqs,Ozsoy:2020ccy}. Combining the resulting expressions with \eqref{br2}, for a given choice of $\delta$, we derive a lower bound on the ratio $f/\Mpl$ in terms of $\epsilon_\phi = r_{\rm v}/16$ \eqref{psv} and $\xi_*$ as
\begin{equation}\label{br2f}
b_0\, e^{b_1 \xi_*}\, \sqrt{r_{\rm v}}  < \frac{f}{M_{\mathrm{pl}}}, 
\end{equation}
where for both models and all the $\delta$ values we consider, the precise values of the coefficients $c_0$ and $c_1$ can be found in Table \ref{tab:br2f}. Comparing the two models we consider from the table, we see that the lower bound is in general less restrictive in the non-compact axion model (M2) compared to the rolling axion model (M1) with the standard cosine potential. Notice that, increasing $\delta$ relaxes the bounds as in this case $\dot{\chi}$ is maximal for a shorter amount of time, reducing width of the gauge field modes that are effected by the roll of $\chi$ and in general the efficiency of particle production. Finally, it should be clear from \eqref{br2f} that, reducing $r_{\rm v}$ relaxes the bound further where the allowed region for $f/\Mpl$ increases.
\smallskip
\begin{table}[t!]
\begin{center}
\begin{tabular}{| c | c | c |}
\hline
\hline
$\{ \rm M1\}$&$b_0 \simeq $&$b_1 \simeq$ \\\hline
$\delta = 0.2$&\scalebox{0.95}{$4.33 \times 10^{-6}$}&\scalebox{0.95}{$2.750$}\\\hline
$\delta = 0.3$&\scalebox{0.95}{$3.31 \times 10^{-6}$}&\scalebox{0.95}{$2.645$}\\\hline
$\delta = 0.4$&\scalebox{0.95}{$ 2.95 \times 10^{-6}$}&\scalebox{0.95}{$2.535$}\\\hline
$\delta = 0.6$&\scalebox{0.95}{$2.79 \times 10^{-6}$}&\scalebox{0.95}{$2.305$}\\\hline
\hline
\end{tabular}\,
\begin{tabular}{| c | c | c |}
\hline
\hline
$\{\rm M2\}$&$b_0 \simeq $ &$b_1 \simeq$ \\\hline
$\delta = 0.2$&\scalebox{0.95}{$3.42 \times 10^{-7}$}&\scalebox{0.95}{$2.745$}\\\hline
$\delta = 0.3$&\scalebox{0.95}{$3.57 \times 10^{-7}$}&\scalebox{0.95}{$2.595$}\\\hline
$\delta = 0.4$&\scalebox{0.95}{$ 3.81 \times 10^{-7}$}&\scalebox{0.95}{$2.445$}\\\hline
$\delta = 0.6$&\scalebox{0.95}{$4.27 \times 10^{-7}$}&\scalebox{0.95}{$2.165$}\\\hline
\hline
\end{tabular}
\caption{\label{tab:br2f} Table of coefficients $c_0, c_1$ that parametrizes the lower bound on $f/\Mpl$ in \eqref{br2f}. }
\end{center}			
\end{table}
\subsubsection{Perturbativity constraints}
The production of scale dependent, chiral GWs of non-vacuum origin in the spectator sector typically demands an exponentially large amplitude in the gauge field sources during the times/at scales when the observable effects are produced. Therefore, one may wonder if large amplitudes obtained by the gauge field fluctuations can drive the system out of the perturbative regime which was the intrinsic assumption we made in deriving the sourced templates of scalar and tensor perturbations in \eqref{SC}. In the following analysis, our aim is therefore to establish the regime for which these results are under perturbative control. For this purpose, we consider two main requirements that the spectator models \eqref{AfM} (and \eqref{pots}) we focus should fulfill \cite{Peloso:2016gqs}:
\begin{enumerate}
\item Higher order loop corrections induced through the interaction \eqref{Lint} do not spoil the leading order estimates for the amplified gauge field mode functions. This criterion can be written in terms of the model parameters as \cite{Peloso:2016gqs,Ozsoy:2020ccy},
\begin{equation}\label{PA}
P_{A}\left(\xi_*,\delta, \frac{k}{k_*},\frac{\tau}{\tau_*}\right) \equiv\left|\frac{\delta^{(1)}\left\langle\hat{A}_{-}(\tau, \vec{k}) \hat{A}_{-}\left(\tau, \vec{k}^{\prime}\right)\right\rangle^{\prime}}{\left\langle\hat{A}_{-}(\tau, \vec{k}) \hat{A}_{-}\left(\tau, \vec{k}^{\prime}\right)\right\rangle^{\prime}}\right| \ll 1,
\end{equation}
where prime denotes the two point function without the corresponding delta function and the expression in the nominator/denominator represent leading order loop correction to the gauge field propagator and the corresponding three level result, respectively.
\item For the second criterion, we will demand that the interaction \eqref{Lint} does not induce a variance $\sqrt{\delta \chi^2}$ that is larger than the typical classical field excursion $\chi_{\rm cl}$. We therefore require \cite{Peloso:2016gqs,Ozsoy:2020ccy},
\begin{equation}\label{Pchi}
P_{\chi}\left(\xi_*,\delta, \frac{\tau}{\tau_*}\right)  \equiv \frac{\sqrt{\left\langle\delta \hat{\chi}^{(1)}(\tau, \vec{x}) \delta \hat{\chi}^{(1)}(\tau, \vec{x})\right\rangle}}{\chi_{\mathrm{cl}}}=\frac{\sqrt{\int \mathrm{d} \ln k\, \mathcal{P}_{\chi}^{(1)}(\tau, k)}}{\chi_{\mathrm{cl}}} \ll 1,
\end{equation}
where we described the numerator as an integral of the leading order loop contribution to the axion's power spectrum $2 \pi^{2} \mathcal{P}_{\chi}^{(1)}(\tau, k) / k^{3}=\left\langle\delta \hat{\chi}^{(1)}(\tau, \vec{k}) \delta \hat{\chi}^{(1)}(\tau,-\vec{k})\right\rangle^{\prime}$. In the first model M1 with standard axion modulations, while it is natural to identify $\chi_{\rm cl} \to f$ \cite{Peloso:2016gqs}, due to non-compact nature of axion in the second model (M2), we will use $\chi_{\rm cl}  \to \chi_{\rm in} \simeq 3\pi f /2$ assuming $\chi$ rolls over one bump like region in its potential before reaching its global minimum at $\chi = 0$ \cite{Ozsoy:2020ccy}.
\end{enumerate}
As indicated by the expression \eqref{PA}, the first criterion is time and scale dependent. To evaluate this expression, we will focus on the mode that is most amplified by the rolling axion, $k = 5 k_*$ by evaluating the expression at a late time $\tau/\tau_* \to 0$ at which the gauge field mode functions are maximally enhanced. As shown in \cite{Peloso:2016gqs,Ozsoy:2020ccy}, this strategy is sufficient to derive strongest constraints as the growth in $P_A$ (and also for $P_\chi$) saturates at late times for the most amplified mode. The second criterion \eqref{Pchi} on the other hand arise as an integral over modes and hence scale independent. Following the procedure outlined in \cite{Peloso:2016gqs,Ozsoy:2020ccy}, we evaluated these criteria for both models we consider and for $\delta = 0.2,0.3,0.4,0.6$ corresponding the increasingly faster rolling axion. Similar to the backreaction constraints we derived earlier, at fixed $\delta$, the resulting constraints can be interpreted as a lower bound on $f/\Mpl$ in terms of $r_{\rm v}$ and the effective coupling $\xi_*$ as
\begin{table}[t!]
\begin{center}
\begin{tabular}{| c | c | c |}
\hline
\hline
$\{\rm M1\}$&$p_0 \simeq $&$p_1 \simeq$ \\\hline
$\delta = 0.2$&\scalebox{0.95}{$2.85 \times 10^{-6}$}&\scalebox{0.95}{$2.820$}\\\hline
$\delta = 0.3$&\scalebox{0.95}{$1.43 \times 10^{-6}$}&\scalebox{0.95}{$2.792$}\\\hline
$\delta = 0.4$&\scalebox{0.95}{$9.96 \times 10^{-7}$}&\scalebox{0.95}{$2.734$}\\\hline
$\delta = 0.6$&\scalebox{0.95}{$8.72 \times 10^{-7}$}&\scalebox{0.95}{$2.549$}\\\hline
\end{tabular}\,
\begin{tabular}{| c | c | c |}
\hline
\hline
$\{ \rm M2\}$&$p_0 \simeq $&$p_1 \simeq$ \\\hline
$\delta = 0.2$& \scalebox{0.95}{$1.66 \times 10^{-7}$}&\scalebox{0.95}{$2.796$}\\\hline
$\delta = 0.3$& \scalebox{0.95}{$1.49 \times 10^{-7}$}&\scalebox{0.95}{$2.710$}\\\hline
$\delta = 0.4$&\scalebox{0.95}{$1.45 \times 10^{-7}$}&\scalebox{0.95}{$2.607$}\\\hline
$\delta = 0.6$&\scalebox{0.95}{$1.80 \times 10^{-7}$}&\scalebox{0.95}{$2.360$}\\\hline
\end{tabular}\\
\smallskip
\begin{tabular}{| c | c | c |}
\hline
$\{ \rm M1\}$&$\tilde{p}_0 \simeq $&$\tilde{p}_1 \simeq$ \\\hline
$\delta = 0.2$&\scalebox{0.95}{$1.24 \times 10^{-6}$}&\scalebox{0.95}{$2.789$}\\\hline
$\delta = 0.3$&\scalebox{0.95}{$1.01 \times 10^{-6}$}&\scalebox{0.95}{$2.698$}\\\hline
$\delta = 0.4$&\scalebox{0.95}{$9.27 \times 10^{-7}$}&\scalebox{0.95}{$2.599$}\\\hline
$\delta = 0.6$&\scalebox{0.95}{$8.94 \times 10^{-7}$}&\scalebox{0.95}{$2.389$}\\\hline
\hline
\end{tabular}\,
\begin{tabular}{| c | c | c |}
\hline
$\{ \rm M2\}$&$\tilde{p}_0 \simeq $&$\tilde{p}_1 \simeq$ \\\hline
$\delta = 0.2$&\scalebox{0.95}{$7.36 \times 10^{-8}$}&\scalebox{0.95}{$2.763$}\\\hline
$\delta = 0.3$&\scalebox{0.95}{$8.88 \times 10^{-8}$}&\scalebox{0.95}{$2.644$}\\\hline
$\delta = 0.4$&\scalebox{0.95}{$1.08 \times 10^{-7}$}&\scalebox{0.95}{$2.507$}\\\hline
$\delta = 0.6$&\scalebox{0.95}{$1.50 \times 10^{-7}$}&\scalebox{0.95}{$2.245$}\\\hline
\hline
\end{tabular}
\caption{\label{tab:pcf} Table of coefficients $p_0,p_1,\tilde{p}_0, \tilde{p}_1$ that parametrizes the lower bound on $f/\Mpl$ in \eqref{pcf}. }
\end{center}			
\end{table}
\begin{align}
\nn P_A \ll 1\,\,\quad &\longrightarrow \quad p_0\, e^{p_1 \xi_*}\, \sqrt{r_{\rm v}}   < \frac{f}{M_{\mathrm{pl}}}, \\ 
\label{pcf} P_\chi \ll 1\,\,\quad &\longrightarrow \quad \tilde{p}_0\, e^{\tilde{p}_1 \xi_*}\, \sqrt{r_{\rm v}}   < \frac{f}{M_{\mathrm{pl}}}, 
\end{align}
where we provide explicit values of the coefficients $p,\tilde{p}$ in Table \ref{tab:pcf}. As can be inferred from the tables, for all the $\delta$ choices we focus, bounds from the renormalization of the gauge field wave functions dominate over the second criterion 2) above, and hence we will ignore the latter hereafter. Another information that we can obtain form Table \ref{tab:pcf} is that constraints on $f/\Mpl$ tend to be weaker in the second model M2 compared to M1.

\smallskip
\noindent{\bf Summary of backreaction and perturbativity constraints.} Focusing on the range $3 < \xi_* < 6.5$ of effective coupling within which interesting phenomenology from spectator axion-gauge field dynamics can arise, we compared the lower bounds obtained on $f/\Mpl$ from backreaction  \eqref{br2f} and perturbativity considerations in \eqref{pcf} using Tables \ref{tab:br2f} and \ref{tab:pcf}. In this way, we found that for $\delta = 0.2$, backreaction constraints dominate over $P_A$ for $\xi_* < 5.97$ in M1 and for all $\xi_*$ range we quoted above in M2. For $\delta = 0.3$, the range of domination of the backreaction constraints reduces to $\xi_* < 5.72$ in M1 while it still dominates over the perturbativity for all $\xi_*$ in M2. Increasing $\delta$ further makes the perturbativity bound stronger compared to backreaction. For example, for the choice of $\delta = 0.4$, backreaction dominates for $\xi_* < 5.45$ in M1 whereas it is stronger than perturbativity for $\xi_* < 6$ in M2. Finally for $\delta = 0.6$, backreaction is stronger than perturbativity for $\xi_* < 4.77$ in M1 while this range is reduced further to $\xi_* < 4.42$ in M2. Considering the upper limits derived in \eqref{br1}, we can then compile all the backreaction and perturbativity constraints on the rolling spectator axion-models as 

\begin{align}\label{sumc}
\nn \quad\quad\textrm{Max}\left[b_0\, \mathrm{e}^{b_1 \xi_{*}} \sqrt{r_{\rm v}},\,\, p_0\, \mathrm{e}^{p_1 \xi_{*}} \sqrt{r_{\rm v}}\right] & < \frac{f}{\Mpl} < \frac{1}{\sqrt{2\delta}},\quad\quad\quad\quad\,\, \textrm{(M1)},\\
\textrm{Max}\left[b_0\, \mathrm{e}^{b_1 \xi_{*}} \sqrt{r_{\rm v}},\,\, p_0\, \mathrm{e}^{p_1 \xi_{*}} \sqrt{r_{\rm v}}\right] & < \frac{f}{\Mpl} <   \frac{1}{\sqrt{5.7\delta}}\quad\quad\quad\quad \textrm{(M2)},
\end{align}
where the coefficients $b,p$ can be read from Tables \ref{tab:br2f} and \ref{tab:pcf} for all $\delta$ choices we focus and following the discussion we presented above.
\begin{figure}[t!]
\begin{center}
\includegraphics[scale=0.62]{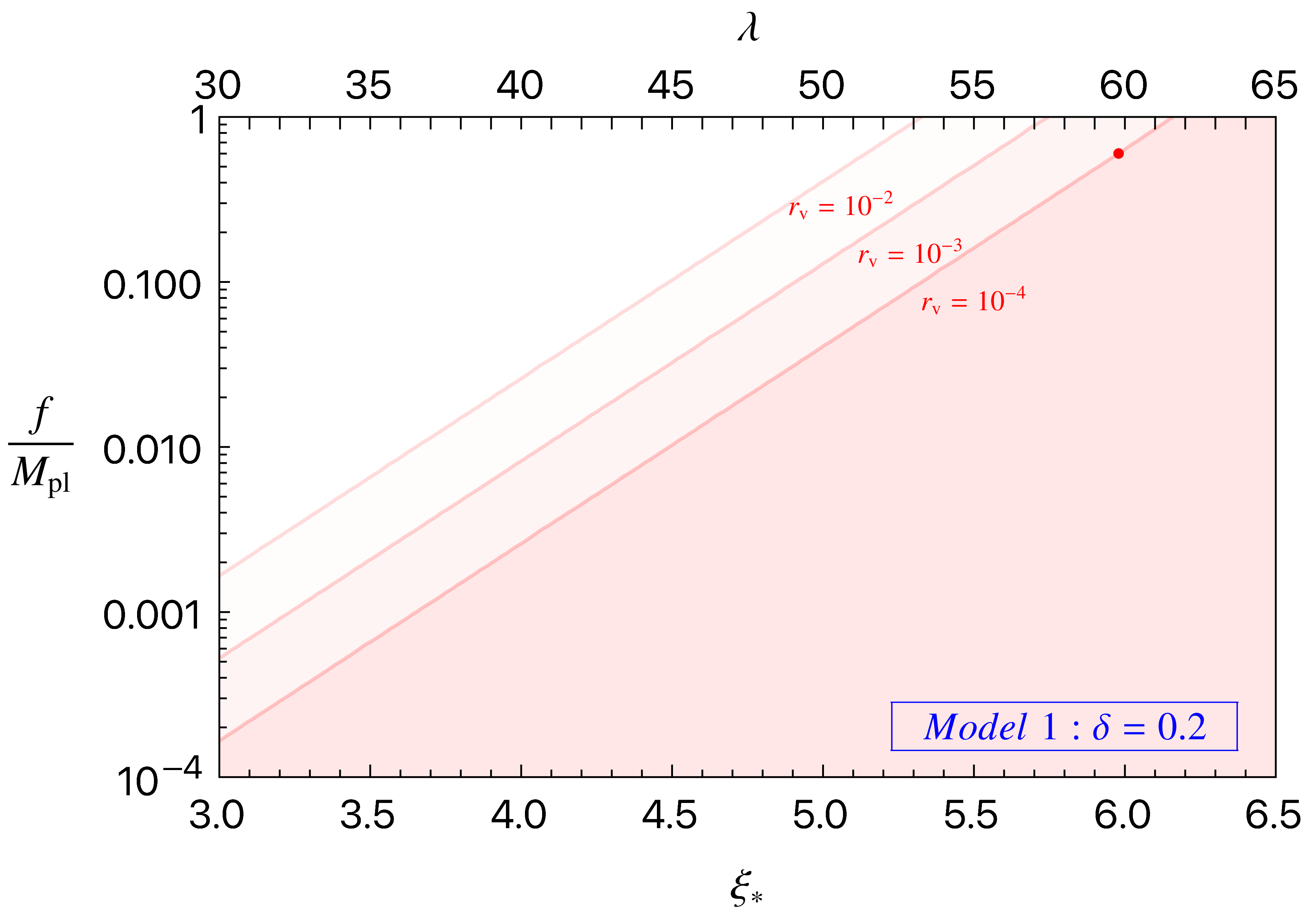}\includegraphics[scale=0.62]{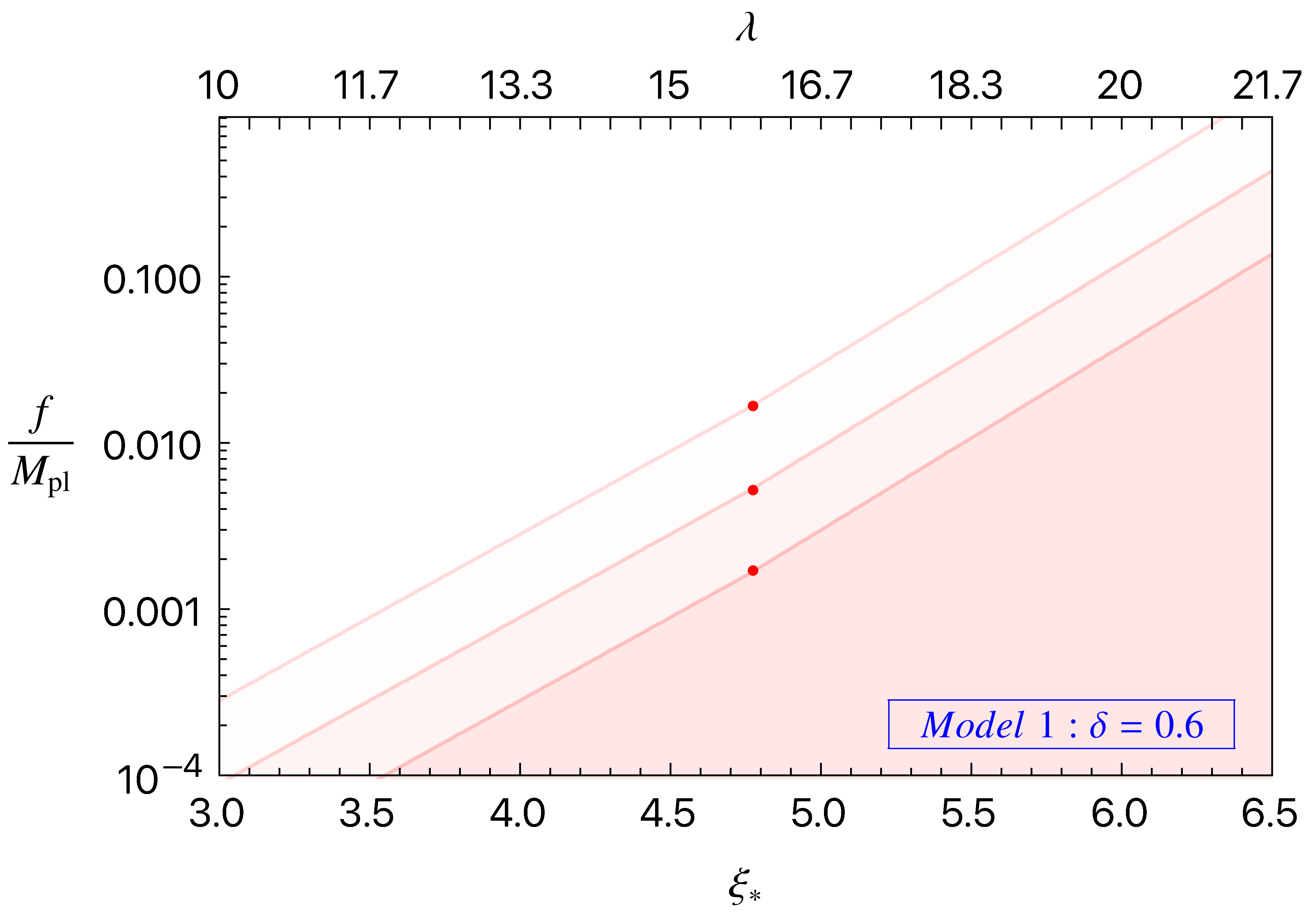}
\includegraphics[scale=0.62]{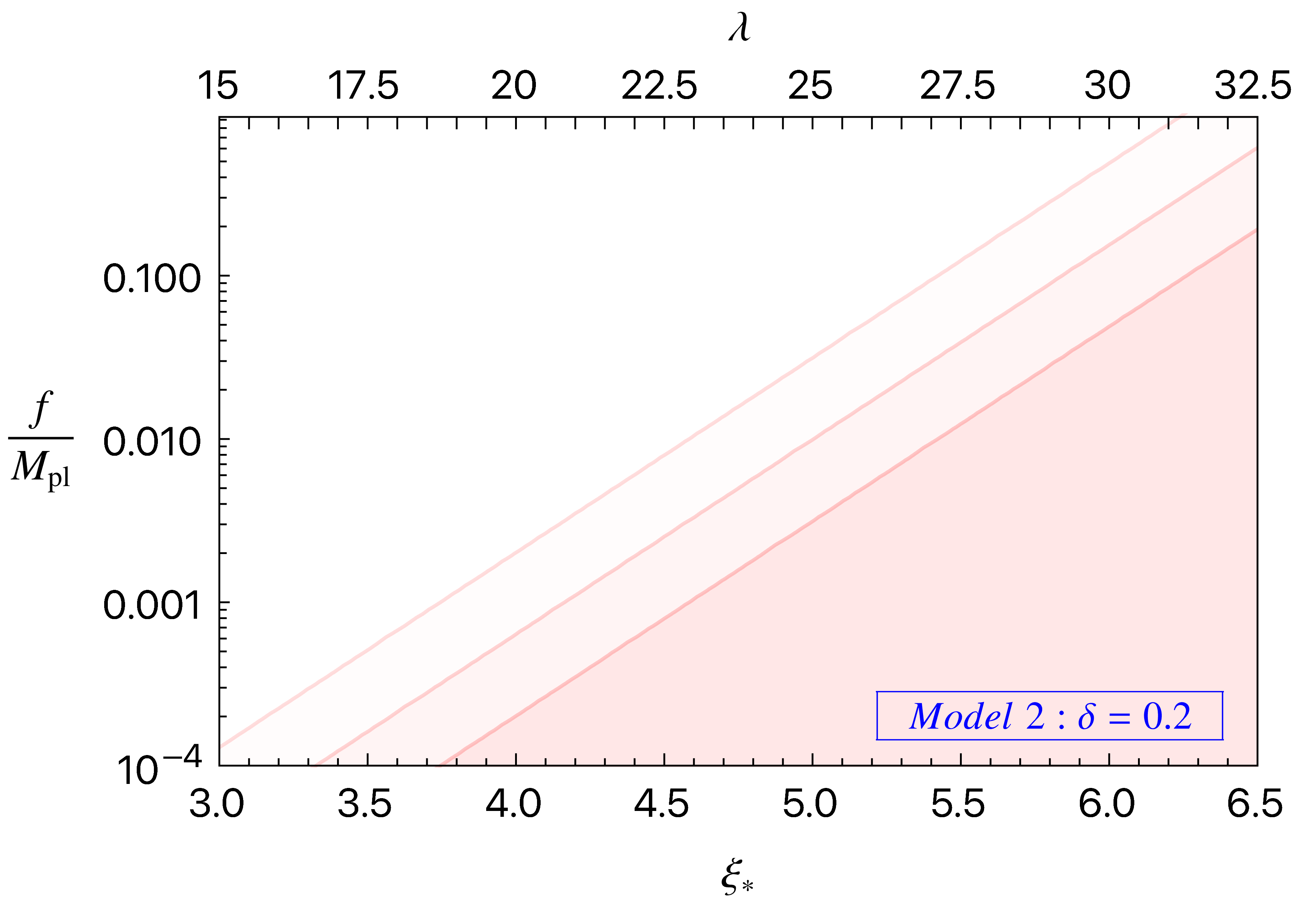}\includegraphics[scale=0.62]{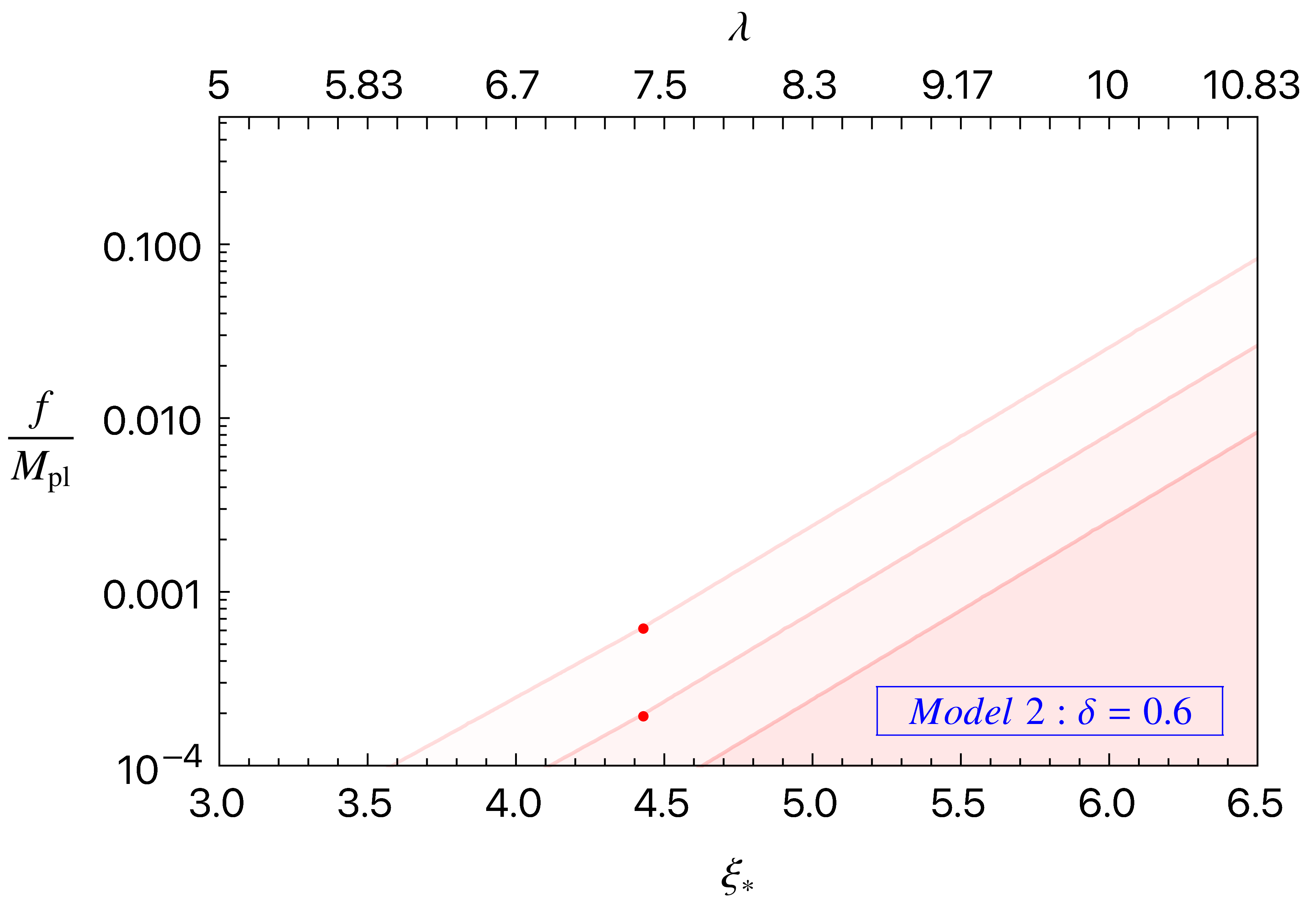}
\end{center}
\caption{The parameter space $f/\Mpl-\xi_*$ ($\lambda$) of the rolling spectator axion models: M1 (top row) and M2 (bottom row), consistent with backreaction and perturbativity bounds (white regions) shown in \eqref{sumc} (see also Tables \ref{tab:br2f} and \ref{tab:pcf}) for $\delta = 0.2$ (Left) and $\delta =0.6$ (Right). The solid lines indicate the lower bound on $f/\Mpl$ for different choices of vacuum tensor-to-scalar ratio: $r_{\rm v} = 10^{-2}$ (top), $r_{\rm v} = 10^{-3}$ (middle) and $r_{\rm v} = 10^{-4}$ (bottom). For each choice of $r_{\rm v}$ shown, the region below these lines (shown by shades of red) violate the aforementioned constraints. The red dots represent the points in the parameter space where the limit on the backreaction becomes comparable with that on the perturbativity in \eqref{sumc}. \label{fig:c}}
\end{figure}

Focusing on the representative cases of $\delta = \{0.2, 0.6\}$, in Fig. \ref{fig:c} we illustrate the parameter space $f/\Mpl - \xi_*$ consistent with perturbativity and backreaction bounds. Note that, using the relation \eqref{Joes} at fixed $\delta$, the same parameter space can be described in terms of the dimensionless coupling $\lambda$ of the spectator axion-gauge field interaction \eqref{Lint} (see the upper x-axis). Comparing the top and bottom panel plots, we confirm that within the same range of $\xi_*$ ($\lambda$) the non-compact axion model M2 has a larger parameter space where backreaction and perturbativity constraints are satisfied. Furthermore, for a faster rolling spectator axion $\chi$ (larger $\delta$), a larger portion of the parameter space opens up for both models\footnote{We confirmed that this conclusion holds for other choices $\delta = \{0.3,0.4\}$. In general, increasing $\delta$, opens up more available parameter space for both models.}. Finally, for smaller $r_{\rm v}$ allowed region for $f/ \Mpl$ enlarges at fixed $\xi_*$ as can be also inferred from the expressions we derived in \eqref{sumc}. The dots indicated by red color in the graphs locate the point in the parameter space where the lower limits derived from perturbativity considerations become comparable to backreaction constraints. In accordance with our discussion above, beyond this point, the dominant constraints on the lower bound for $f/\Mpl$ come from $P_A$ in \eqref{pcf}, resulting with a slight change in the slope of the limiting line (see $p_1$ vs $b_1$ from Tables \ref{tab:br2f} and \ref{tab:pcf}). This change in the slope is barely recognizable from Fig. \ref{fig:c} since we are showing the constraints using a linear-log scaling.
\newpage
To sum up our findings in this section, the energy density contained in the spectator axion sector is approximately given by $\rho_\chi \approx H^2 f^2$ (see eq. \eqref{br1f}). Therefore, the requirement \eqref{br1} on the sub-dominance of axion's energy density with respect to the inflaton sector, $\rho_\phi \simeq 3H^2 \Mpl^2$, give rise to an upper bound \eqref{br1f} on the scale $f$ with respect to $\Mpl$. On the other hand, backreaction and perturbativity criteria we discussed above can be translated into a lower bound on the scale $f$ (see eqs. \eqref{br2f} and \eqref{pcf}). This is because lowering the scale $f$ with respect to $\Mpl$ leads to a stronger interaction \eqref{Lint} between spectator axion and ${\rm U}(1)$ gauge fields, increasing the efficiency of the particle production in the gauge field sector. At fixed $\delta$ and vacuum tensor-to-scalar ratio $r_{\rm v}$, we showed that this lower bound is dictated by the effective coupling $\xi_*$ ($\lambda$) between the spectator Abelian gauge and axion field and parametrized by the first inequality in eq. \eqref{sumc}.

\section{New constraints on scale-dependent GWs sourced by vector fields from \textit{Planck} and BICEP/Keck data}
\label{sec:observation}

While the previous section was dedicated to the evaluation of theoretical bounds given by the backreaction and perturbativity considerations, we will devote this section to provide observational constraints on the parameter space of the spectator axion-${\rm U}(1)$ gauge field model \eqref{AfM}, and specifically on the transiently rolling axion models (see Section \ref{rams}) M1 and M2 parametrized by the scalar  potentials in \eqref{pots}. 

\smallskip
\noindent{\bf CMB 2-point function analysis.}
For this purpose, we use the latest \textit{Planck} \texttt{NPIPE}-processed PR4 maps release \cite{Tristram:2020wbi} and BICEP/Keck 2018 (hereafter BK18) data \cite{BICEP_2021}, which together represent the state-of-the-art dataset for constraining primordial scalar  fluctuations and tensor modes at the largest cosmological scales we are interested in this paper\footnote{Note that in principle, the M2 model can also simultaneously produce sizeable amount of GWs at sub-CMB scales, which can be probed for instance through pulsar timing arrays and laser interferometers \cite{Ozsoy:2020ccy}.}. Similarly to analysis carried in \cite{Namba:2015gja} using a simplified data analysis setup (i.e. by fixing all model's, cosmological and likelihood nuisance parameters except for the $\xi_{*}$ parameter) with WMAP temperature data, by exploiting the full power of present day CMB temperature and polarization datasets, we aim to show the extent to which the models considered in this work can produce an observable amount of sourced gravitational waves, while remaining consistent with current tight constraints on the scalar sector. 

In the following, we perform a likelihood analysis leaving the $\xi_{*}$ parameter free (together with cosmological and likelihood nuisance parameters) and fixing $\delta$ and $k_{*}$. As motivated in Section \ref{rams}, for $\delta$,  we consider the following representative values
\begin{equation}
 \delta = \{0.2,\, 0.3,\, 0.4,\, 0.6\},  \nonumber 
\end{equation} 
ordered according to decreasing amplitude (at fixed $\xi_{*}$) and increasing sharpness of the sourced Gaussian bump \eqref{SC}.  For $k_{*}$, we choose instead the  following three representative values:
\begin{equation}\label{kstar}
    k_{*} = \{\, 7\times 10^{-5},\, 5\times 10^{-4},\, 5\times 10^{-3}\,\}\,[{\rm Mpc}^{-1}]. \nonumber
\end{equation}
For the models under consideration, the first value above typically induces a sourced bump in the scalar and tensor spectra at the very largest CMB scales (pertaining to the reionization bump in $E$ and $B$-modes spectra), the second one affects the recombination bump's multipole range in $B$-modes, and finally the third one impacts scales around the first acoustic peak again in $B$-modes. Note from eq. \eqref{SC} that the actual scale at which the sourced bump appears is different from the critical scale $k_{*}$: $k_{\rm p} = k_{*}\, x^{c}_{2,-}[\delta, \xi_{*}] > k_*$ because of the momentum conservation law of one-loop interactions that generates the sourced signals\footnote{In particular, at the time when axion's velocity peaks ($\tau_* = k_*^{-1}$), maximally amplified gauge field modes (running in the loop) are typically inside the horizon obeying $q > k_* \mathcal{O}(1)$ \cite{Namba:2015gja,Ozsoy:2020ccy}. Due to momentum conservation at each vertex ($2 \times (A_i + A_i \to \{h_\lambda,\mathcal{R}\})$) contributing to the one-loop power spectra, the resulting correlations among the external states ($\{h_\lambda,\mathcal{R}\}$) is therefore maximal for wave-numbers satisfying $k = k_{\rm p} > k_{*}$.}.
This deviation becomes larger at fixed $\xi_{*}$ for increasing $\delta$, as can be inferred from Tables \ref{tab:fit1}-\ref{tab:fit4}. Furthermore, $k_{\rm p}$ is typically larger for the model M2 compared to M1 (at fixed $k_{*}$, $\xi_{*}$ and $\delta$): we will see in the following that this has important consequences on the model constraints.

\smallskip
\noindent{\bf CMB 3-point function and parity-violating correlations.}
As highlighted in the previous literature \cite{Namba:2015gja, Shiraishi:2019yux}, the strict constraints on scalar non-Gaussianity at smaller scales can be evaded in the models under examination, if we consider a sourced bump at large scales in the spectra, generated by an axion rolling for only a few e-folds $\Delta N_{\chi} \sim \delta^{-1}$ during inflation. For example, this condition is realized if the axion velocity satisfies $\dot{\chi}\to 0$ (or $\xi \to 0$ as in \eqref{Joep}) when modes with $\ell \gtrsim \mathcal{O}(10^2)$ leave the horizon so that gauge field production is ineffective at those scales. Tensor  non-Gaussianity\footnote{See also \cite{Ozsoy:2021onx} for the study of mixed scalar-tensor type non-Gaussianity in the spectator axion-gauge field models we consider in this work.} \cite{Shiraishi:2016yun} generated in spectator axion-${\rm U}(1)$ model can also provide complementary information to CMB 2-point functions, even though the most stringent constraints are still obtained from the latter. The sourced bump in $BB$ is indeed accompanied by a similar bump in the $BBB$ 3-point function, which however has smaller signal-to-noise ratio compared to $BB$ one \cite{Shiraishi:2019yux}. Therefore, the analysis in the following will be based solely on CMB 2-point functions, and we leave a thorough analysis including bispectrum constraints to future work.
Finally, as we discussed in Section \ref{rams}, the models we consider produce fully chiral gravitational waves: the possibility of detecting such circular polarization with the CMB has been addressed in previous literature \cite{Namba:2015gja, Gerbino:2016mqb}. However, parity-violating $EB$ and $TB$ correlations in \textit{Planck} data can constrain only very weakly the chirality parameter \cite{Gerbino:2016mqb}, therefore we will not consider them further in our analysis.

\subsection{Data and likelihoods}
As anticipated above, we exploit the latest \textit{Planck} and BK18 public data releases, with likelihoods publicly available for the $\texttt{Cobaya}$ \cite{Torrado:2020dgo} MCMC framework. Specifically, we combine the low-$\ell$ $TT$ \texttt{Commander} likelihood (covering multipoles $\ell=2-30$) with the high-$\ell$ $TT+TE+EE$ \texttt{HiLLiPoP} likelihood in the range $\ell = 30 - 2500$ and the low-$\ell$ ($\ell = 2 -150$) $EE+BB+EB$ \texttt{LoLLiPoP} likelihood, as described in \cite{Tristram:2020wbi}. In the analysis, we also include the $B$-mode intermediate-scale constraints from BK18, neglecting correlations with \textit{Planck}. This is doable because $B$-modes are noise-dominated and the two CMB surveys have uncorrelated noises and, moreover, they observe very different fractions of sky \cite{Tristram:2020wbi, Tristram:2021tvh}. 

In the following, we will also study the impact of separate datasets on the constraints: in particular, we name \textit{Planck TT} the combination of low-$\ell$ $TT$ \texttt{Commander} and  \texttt{HiLLiPoP} $TT$ likelihoods, while \textit{Planck TEB} is the combination of \textit{Planck TT}, \texttt{HiLLiPoP} $TE +EE$ and \texttt{LoLLiPoP} $EE$, $BB$ and $EB$ likelihoods.

\subsection{Methodology: the profile likelihood}
In order to provide constraints on the model parameters, we perform a frequentist profile likelihood analysis. Compared to the Bayesian framework widely used in cosmology, frequentist methods have been applied in fewer occasions, despite having several advantages \cite{Cousins:1994yw, planck_profile}. First, they do not require to choose arbitrary priors on the parameters, a practice which may have an important impact on the final bounds in a Bayesian setting. Second, the maximum likelihood estimate (MLE) is invariant under different choices of the model parameterization. Third, frequentist parameter estimates are not affected by so-called ``volume effects'' \cite{Hamann:2007pi}, which can instead appear, due to the marginalization process, in Monte Carlo Markov Chain analysis\footnote{Volume effects arise because marginalization enhances regions of the parameter space that contain more probability density volume in the marginalised directions. Moreover the volume of probability density in a certain parameter direction depends on both the choice of priors and the model parameterization. This can often result into the peak of the marginalized posterior distribution being far from the global MLE.}.

\begin{figure}
    \centering
    \includegraphics[scale=0.52]{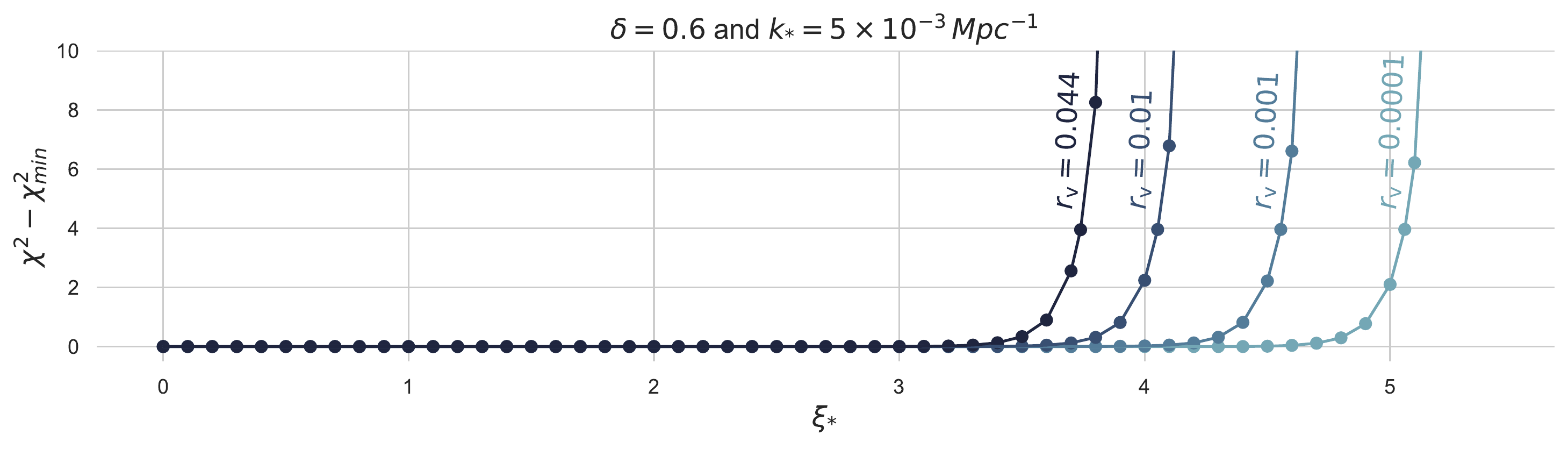}
    \caption{Profile likelihood for $\xi_{*}$. The example shown assumes the M1 model, $\delta=0.6$ and $k_{*}=5\times 10^{-3}\, {\rm Mpc}^{-1}$ with and different colors corresponding to different values of $r_{\rm v}$.}
    \label{fig:profile_example}
\end{figure}

The profile likelihood for a given parameter of interest $\theta$ is obtained by fixing $\theta$ to  a certain value within the range of interest and maximizing the likelihood with respect to all remaining parameters. The maximisation is then repeated for several different values of the parameter of interest, scanning a wide range of $\theta$ values. The minimum of the profile likelihood built in this way coincides, by construction, with the global MLE given the full parameters set. Specifically, we minimize the $\chi^{2}$ function using the \texttt{iMinuit} multi-dimensional minimizer package \cite{iminuit}, a \texttt{python} implementation of the famous \texttt{Minuit} algorithm \cite{Minuit}.
A typical example of profile likelihood for our parameter of interest $\xi_{*}$ is shown in Fig. \ref{fig:profile_example} for the M1 model where we focus on $\delta=0.6$ and $k_{*}=5\times 10^{-3}\,{\rm Mpc}^{-1}$ for four representative values of the vacuum tensor-to-scalar ratio\footnote{We also attempted building the profile likelihood by fitting $r_{\rm v}$ in addition to $\xi_{*}$, the cosmological and the nuisance parameters. However, because of the degeneracy between $r_{\rm v}=16\epsilon_{\phi}$ and $\xi_{*}$ \eqref{SC}, the latter remains essentially unconstrained when fitted together with $r_{\rm v}$. This happens because both $\xi_{*}$ and $\epsilon_{\phi}$ control the amplitude of the sourced signals, so it is always possible to decrease $r_{\rm v}$ to accommodate for larger $\xi_*$. Therefore, we fixed $r_{\rm v}$ to phenomenologically reasonable values in order to obtain more informative constraints.}
\begin{equation}
    r_{\rm v}=\{0.0001,\, 0.001,\, 0.01,\, 0.044\}.\nonumber
\end{equation}
The behaviour of the profile likelihood reflects the exponential nature of the gauge field production \cite{Anber:2009ua}, with a steep growth starting at increasing $\xi_{*}$ value for decreasing $r_{\rm v}$. $\Delta\chi^2$ is instead zero for smaller values of $\xi_{*}$, since the amount of sourced modes produced by the model is negligible and does not affect the likelihood.

An upper bound on $\xi_{*}\equiv\xi_{*,limit}$ is obtained by cutting the profile likelihood $\Delta\chi^2=\chi^2 - \chi^2_{min}=4$ in each of the cases considered. 
We note that the limits derived in this paper cannot be directly compared to the ones derived in \cite{Namba:2015gja}, because in the latter all parameters were fixed to WMAP $\Lambda$CDM best-fit values except $\xi_{*}$. Therefore the approach in \cite{Namba:2015gja} is not guaranteed to reach the global MLE of the likelihood, while in the profile likelihood approach used in this paper we vary all parameters (model + cosmological + nuisance) and the result matches the global minimum of the likelihood up to numerical accuracy\footnote{We also checked that the minimizer was not trapped in any local minimum, by starting minimization from a wide range of different initial parameters sets.}.

\subsection{Observational bounds from the CMB: results and discussion}\label{3p3}
\begin{figure}
    \centering
    \includegraphics[scale=0.6]{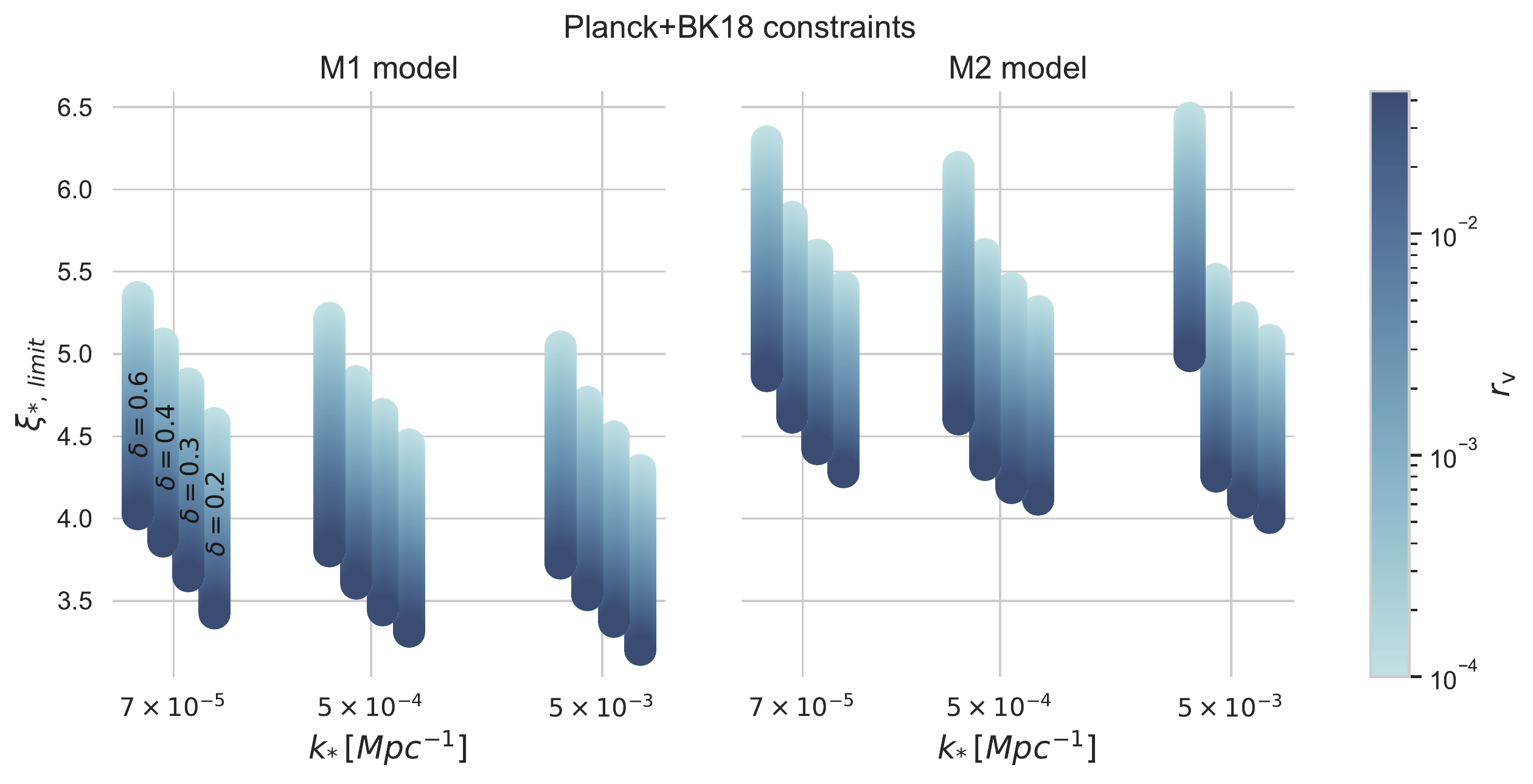}
    \caption{Upper limit on $\xi_{*}$ from \textit{Planck} + BK18 data for the M1 (left panel) and M2 (right panel) models. For each of the three $k_{*}$ values on the $x$-axis, the four bars represents a different value of $\delta$ in decreasing order, colored according to the value of $r_{\rm v}$.}
    \label{fig:upper_xi_vs_k}
\end{figure}

\begin{figure}
    \centering
    \includegraphics[scale=0.6]{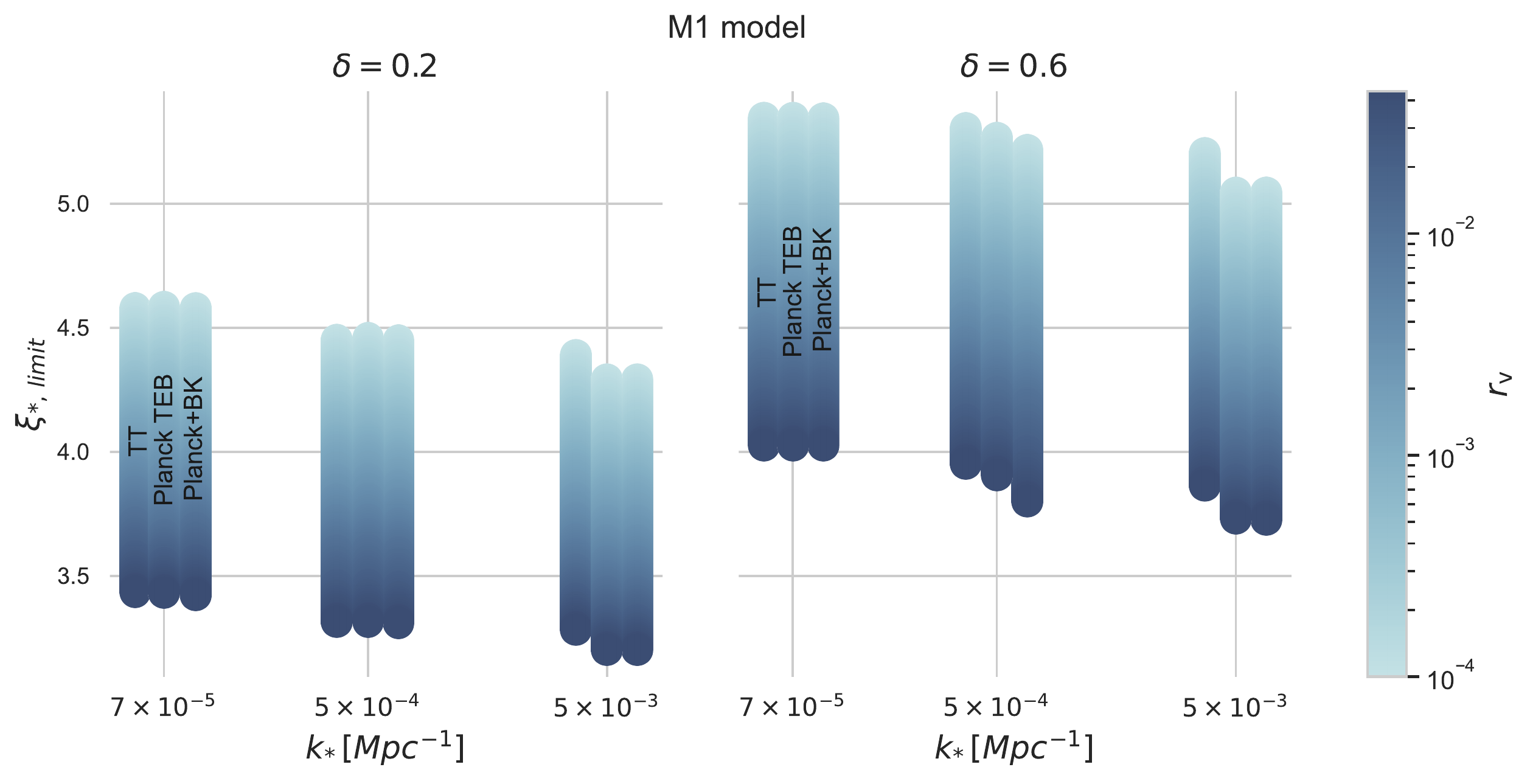}
    \caption{Comparison of upper limit on $\xi_{*}$ for different datasets (\textit{Planck TT}, \textit{Planck TEB} and \textit{Planck} + BK18), for three values of $k_{*}$ and as a function of $r_{\rm v}$. The left panel assumes $\delta=0.2$, while the right one $\delta=0.6$. We assume here the M1 model.}
    \label{fig:TT_vs_TEB}
\end{figure}

\begin{figure}
    \centering
    \includegraphics[scale=0.6]{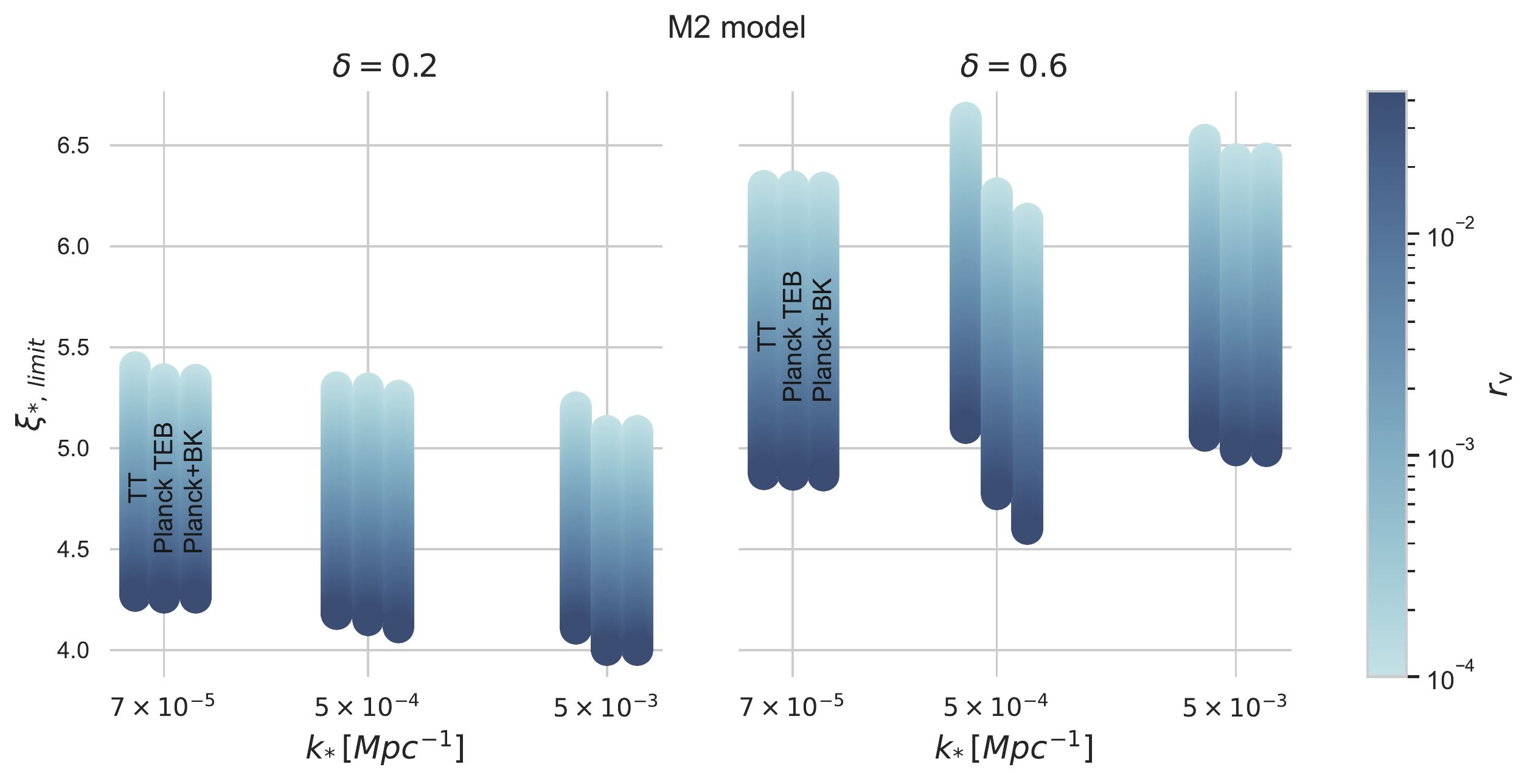}
    \caption{Same as previous figure but assuming the M2 model.}
    \label{fig:TT_vs_TEB_M2}
\end{figure}

\begin{figure}
    \centering
    \includegraphics[scale=0.6]{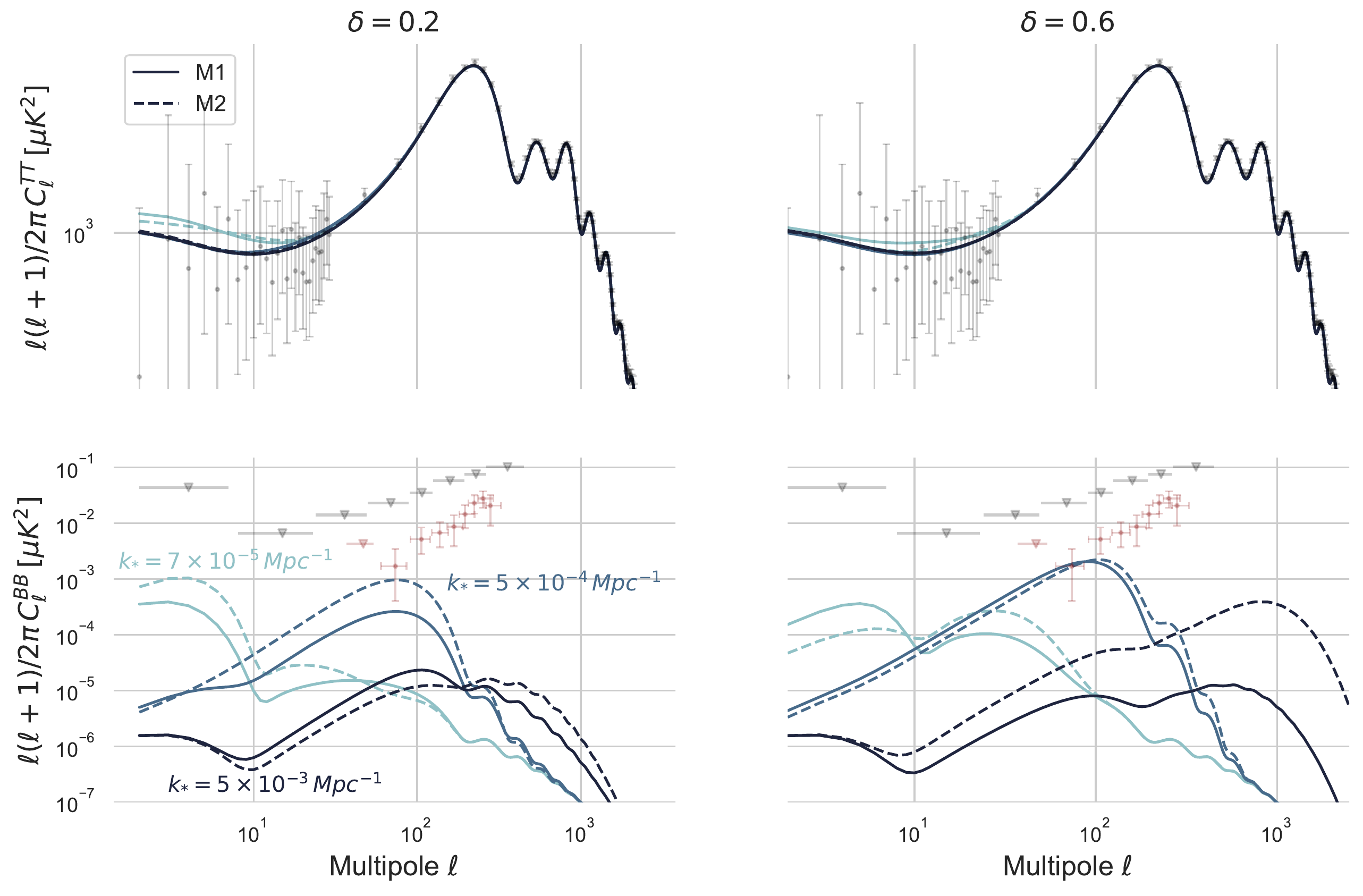}
    \caption{Theoretical CMB spectra evaluated at $\xi_{*,\, limit}$ for \textit{Planck} + BK18. Plots in the left (right) column assume $\delta=0.2$ ($\delta=0.6$). The M1 (M2) model is shown by solid (dashed) lines in three different colors corresponding to the three values of $k_{*}$ considered. Here we assume $r_{\rm v}=10^{-4}$. We also report as reference the 95\% C.L. error bars and upper limits from \textit{Planck} (in gray) and BK18 (in red) data.}
    \label{fig:Cls_theory}
\end{figure}

\begin{figure}
    \centering
    \includegraphics[scale=0.524]{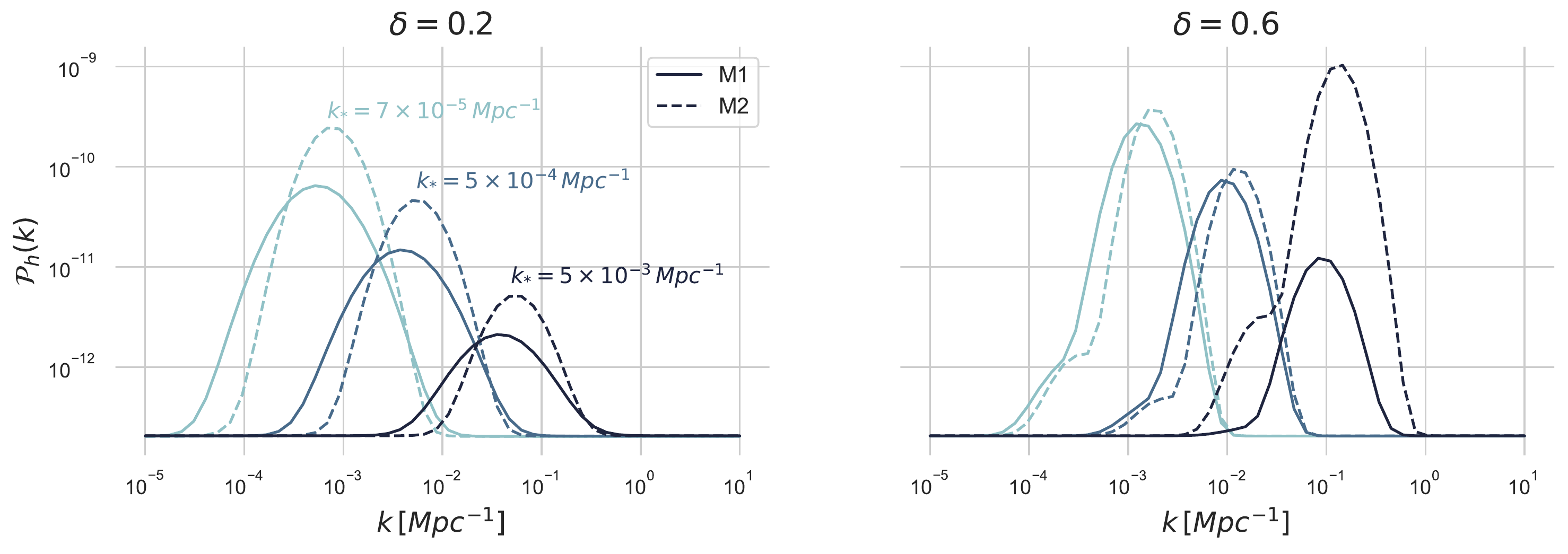}
    \caption{Total (vacuum + sourced) primordial tensor power spectra $\mathcal{P}_{h}(k)$ evaluated at $\xi_{*,\, limit}$ for \textit{Planck}+BK18. The left (right) plot assumes $\delta=0.2$ ($\delta=0.6$). The M1 (M2) model is shown by solid (dashed) lines in three different colors corresponding to the three values of $k_{*}$ considered. Here we assume $r_{\rm v}=10^{-4}$.}
    \label{fig:PPS_theory}
\end{figure}

The upper bounds on the $\xi_{*}$ parameter, obtained from the latest \textit{Planck} and BK18 datasets, are summarized in Fig. \ref{fig:upper_xi_vs_k} for both the M1 and M2 models. The impact of separate datasets (i.e. \textit{Planck TT}, \textit{Planck TEB} and \textit{Planck} + BK18) on the constraints is singled out in Fig. \ref{fig:TT_vs_TEB} for M1 and Fig. \ref{fig:TT_vs_TEB_M2} for M2, considering the two representative values $\delta=0.2, 0.6$. Finally, in Figures \ref{fig:Cls_theory} and \ref{fig:PPS_theory}, respectively, we show the theoretical CMB spectra and the total (vacuum + sourced) primordial tensor power spectra $\mathcal{P}_h(k)$ evaluated at $\xi_{*,\,limit}$ for some representative cases\footnote{The theoretical CMB spectra are evaluated at the best-fit cosmological parameters obtained by likelihood minimization at fixed $\xi_{*}=\xi_{*,\,limit}$.}. We discuss the bounds for each model separately below, starting from the M1 model.

\smallskip
\noindent{\bf Bounds on M1 model.}
For every $\delta$ considered, the upper limit on $\xi_{*}$ becomes tighter as the sourced bump moves from larger to smaller scales (Fig. \ref{fig:upper_xi_vs_k}, left panel): the reason is that most of the constraining power is coming from scalar modes sourced in the $TT$ and $EE$ spectra. \textit{Planck} large-scale $TT$ modes are indeed cosmic variance-limited, so the trend can be imputed mainly to decreasing cosmic variance at smaller scales. 
 Also, $\xi_{*,\,limit}$ is tighter for smaller $\delta$ at fixed $k_{*}$: the width of the sourced bump is indeed proportional to $1/\delta$, as sourced modes are produced only while the axion is significantly rolling (i.e. for a number of e-folds $\Delta N_\chi \simeq 1/\delta$ \cite{Namba:2015gja}). The sharper the bump is, the fewer multipoles are affected, and so the constraints on $\xi_{*}$ will be generally weaker. In addition, decreasing $\Delta N_\chi$ reduces more the production of sourced scalars than that of tensors: the process $\delta A + \delta A \to \delta \chi \to \delta \phi$ is indeed very sensitive to the axion's velocity \cite{Namba:2015gja} and therefore to $\Delta N_\chi$.      

We now compare constraints from \textit{Planck TT}, \textit{Planck TEB} and \textit{Planck} + BK18. Figure \ref{fig:TT_vs_TEB} confirms that $\xi_{*,\,limit}$ is governed by temperature data: limits do not improve significantly when adding \textit{Planck} $E$ and $B$-modes, except when the bump is sourced at the first acoustic peak scales (i.e. for $k_{*} = 5\times 10^{-3}\, {\rm Mpc}^{-1}$). In this case, since the addition of BK18 $B$-mode data has no significant effect, we conclude that \textit{Planck} intermediate/small scale $E$ modes are providing the extra constraining power.   
Adding BK18 data has no effect for wider bumps, but can slightly tighten the upper bound in the case $\delta=0.6$, when sourced tensors are produced around recombination bump scales, since BK18 is sensitive only to multipoles $\ell \simeq 30 - 250$ (Fig. \ref{fig:Cls_theory}). 

\newpage
\smallskip
\noindent{\bf Bounds on M2 model.}
Let us now discuss the constraints on the M2 model (Figure \ref{fig:upper_xi_vs_k}, right panel). The $\xi_{*,\,limit}$ allowed for M2 is larger than the one for M1 in all cases considered:  the axion's velocity profile in the M2 model \eqref{Joep} is indeed sharper than the M1 one (see also Fig. \ref{fig:PPS_theory}), and therefore allows for larger sourced tensors production for the same level of sourced scalars \cite{Ozsoy:2020ccy}.   
While in the M1 case $\xi_{*,\,limit}$ is always tighter at larger $k_{*}$ and larger $\delta$, for M2 this holds only for $\delta<0.6$: similarly to the M1 model, scalars drive the constraints for wider bumps, while for $\delta=0.6$ tensors play a crucial role. For a bump at the largest and smallest scales considered (i.e. $k_{*}= 7\times 10^{-5}$ and $5\times 10^{-3}\, {\rm Mpc}^{-1}$), indeed, constraints do not improve significantly when adding polarization data (Fig. \ref{fig:TT_vs_TEB_M2}), and are again primarily driven by temperature data. At intermediate scales ($k_{*}= 5\times 10^{-4}\, {\rm Mpc}^{-1}$), instead, constraints substantially improve when adding \textit{Planck} $E$ and $B$ data and even more when adding BK18 data, confirming that tensor modes are driving the constraints.
Furthermore, the peak of the signal moves to larger $k_{\rm p}$ for M2 compared to M1 at fixed $k_{*}$, as can be seen in Fig. \ref{fig:Cls_theory} and \ref{fig:PPS_theory}. This is exacerbated at large $\delta$ and contributes to the loose $\xi_{*,\,limit}$ allowed for $\delta=0.6$.

To summarize our findings, when the bump is sourced at recombination bump scales or smaller, adding polarization enhances the constraining power on both models considered. Moreover, the difference between the \textit{Planck TT} and \textit{Planck TEB} upper limits at fixed $k_{*}$ is larger for larger $\delta$, since for a sharper bump the scalar constraints allow for a larger value of $\xi_{*}$, increasing the SNR in polarization (especially in $BB$) and therefore making polarization data more relevant. On the other hand, for a sourced signal peaking at the largest scales considered, adding polarization data has minor or no impact on both models constraints, compatibly with the low sensitivity of \textit{Planck} $B$-modes at the very largest scales \cite{Tristram:2020wbi, Tristram:2021tvh}. 
\newpage

\begin{figure}
    \centering
    \includegraphics[scale=0.62]{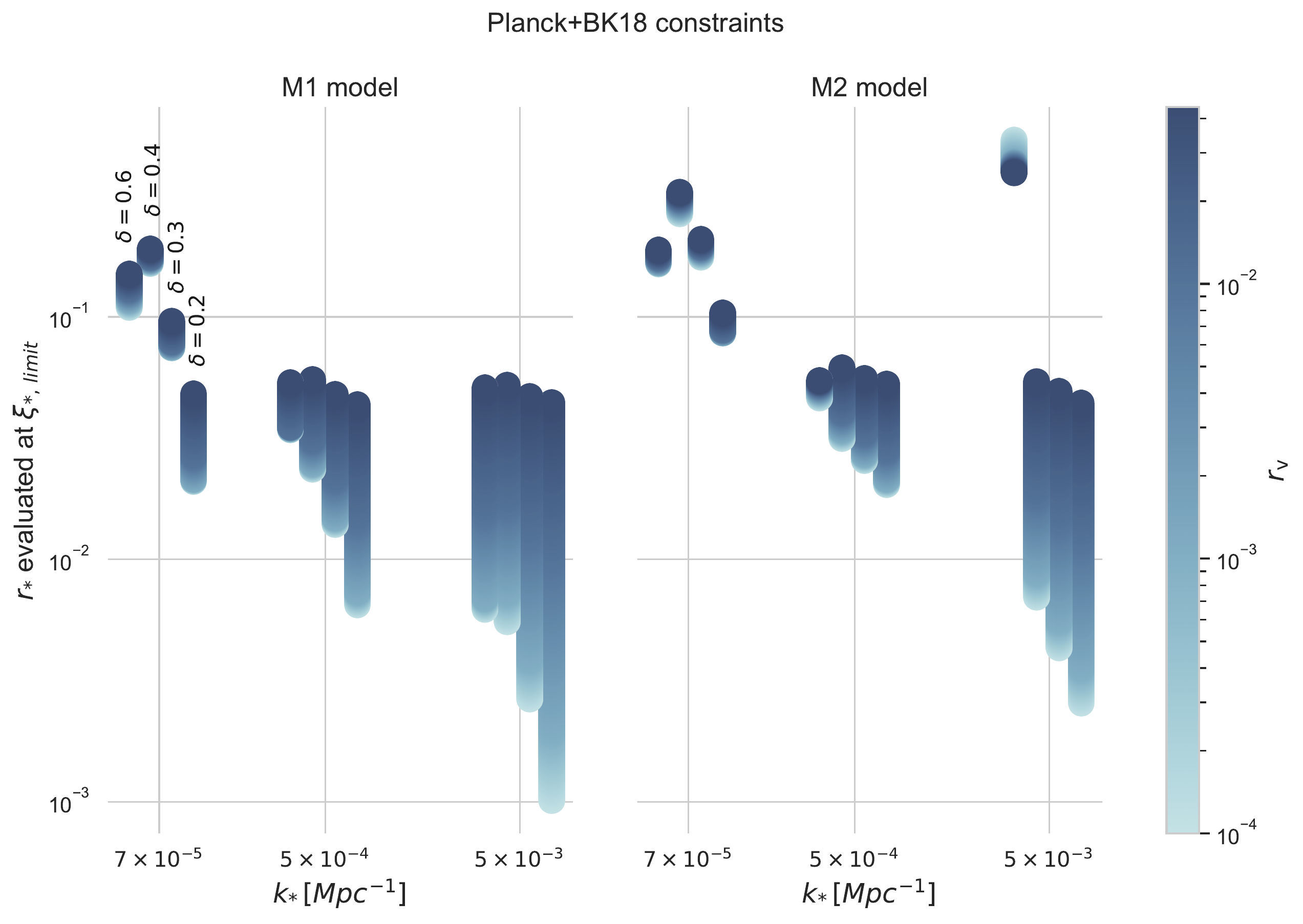}
    \caption{Total (vacuum + sourced) tensor-to-scalar ratio $r_{*}(k_{\rm p})$ evaluated at $\xi_{*,\, limit}$ for the M1 (left panel) and M2 (right panel) models. All spectra are evaluated at the peak of the sourced signal $k_{\rm p} = k_{*}\, x^{c}_{2,-}[\delta, \xi_{*}]$. }
    \label{fig:rstar}
\end{figure}

\smallskip
\noindent{\bf Total tensor-to-scalar ratio.}
Finally, in Fig. \ref{fig:rstar}, we show the values of the total (i.e. vacuum + sourced) tensor-to-scalar ratio $r_{*}(k)$ \cite{Namba:2015gja},
\begin{equation}\label{r_star}
    r_{*}(k) = \frac{\sum_{\lambda = \pm}  \left[
    \mathcal{P}^{(\rm v)}_{\lambda}(k) + \mathcal{P}^{(\rm s)}_{\lambda}(k) 
    \right]
    }{\mathcal{P}^{(\rm v)}_{\mathcal{R}}(k) + \mathcal{P}^{(\rm s)}_{\mathcal{R}}(k)}
\end{equation}
evaluated at $\xi_{*,\,limit}$ for both models. All spectra appearing in \eqref{r_star} are evaluated at the peak of the sourced signal $k = k_{\rm p} = k_{*}\, x^{c}_{2,-}[\delta, \xi_{*}]$. 
Compatibly with our previous discussion, sourced signals peaking at the largest scales generally allow for larger $r_{*}$ values\footnote{An anomalously large value of $r_{*}$ (especially when compared to other $\delta$ cases at the same $k_{*}$) is allowed in the M2 model for $\delta=0.6$ and $k_{*} =  5\times 10^{-3}\, {\rm Mpc}^{-1}$. This happens because, for such high $\delta$, the signal is sharply peaked around $k_{\rm p}$ and, furthermore, $k_{\rm p}$ is very large at such high $k_*$ and $\delta$ (i.e. $k_{\rm p} \simeq 0.14 \,{\rm Mpc}^{-1}$ for $\delta=0.6$ while $k_{\rm p} \lesssim 0.09\,{\rm Mpc}^{-1}$ for the other $\delta$ values). At such small scales, \textit{Planck} and BK18 have essentially no constraining power, resulting in very large allowed $r_{*}$.}. 
Interestingly, the $\delta=0.4$ case allows for the highest $r_*$ at the largest scales for both models: this is because it represents a good compromise between a signal not so spiky that it cannot compensate for the correct normalization of the total scalar power spectrum with sourced modes, and one that is spiky enough that it allows for large values of $\xi_{*}$. 
The production of a sizeable amount of sourced tensor modes, while still complying with scalar constraints, is thus realized.
Moreover, Fig. \ref{fig:rstar} highlights the fact that it is still possible to get significant contribution to $r_{*}$ from sourced modes (reaching $r_{*}\sim \mathcal{O}(10^{-2})$) in the $k_{*} = 5\times 10^{-4}\,{\rm Mpc}^{-1}$ case, even with a vacuum contribution as small as $r_{\rm v}= 10^{-4}$ or $10^{-3}$. On the other hand for $k_{*} = 5\times 10^{-3}\,{\rm Mpc}^{-1}$, the allowed sourced contribution is smaller but can still be significant, especially for the second model M2.

\smallskip
\noindent{\bf Summary of observational constraints.}
To summarize the content of this section,  using the latest \textit{Planck} and BK18 CMB datasets, we derived, for the first time in the literature, constraints on the effective coupling parameter $\xi_{*}$ of the spectator axion-${U}(1)$ gauge field model in \eqref{AfM} for two possible potentials of the transiently rolling spectator axion in \eqref{pots}. We used a fully frequentist profile likelihood approach to derive upper bounds on $\xi_{*}$ which are independent from prior distributions and model parametrization choices and thus immune to volume effects. We provided a detailed interpretation of the behaviour of the upper bound $\xi_{*,\,limit}$ for different choices of the $\delta$ and $k_{*}$ model parameters and compared these results for the two axion potentials under consideration. 
In conclusion, as can be seen by comparing Figures \ref{fig:c} and \ref{fig:upper_xi_vs_k}, the observational bounds reported in this section are competitive with the theoretical bounds from perturbativity and backreaction (Section \ref{sec:theory}). We will address in the next section the effect of these combined theoretical and observational bounds on the model's parameter space, in the context of particle production in the spectator sector.

\section{Conclusions}\label{sec:conclusions}

At the end of this decade, new CMB probes, such as the LiteBIRD satellite \cite{Hazumi:2019lys} and the ground-based CMB-S4 \cite{CMB-S4:2016ple}, will target the imprint in the $B$-mode polarization pattern left by the primordial gravitational waves. In case of a detection, however, it will still be necessary to perform further tests in order to understand the origin of this signal and distinguish between the SGWB generated by quantum vacuum fluctuations of the metric, within the leading paradigm of single-field, slow-roll inflation, and the one possibly sourced by additional matter fields present during inflation. The SGWB properties predicted in these two scenarios can greatly differ, e.g. an almost scale-invariant spectrum from quantum vacuum fluctuations versus a strongly scale-dependent one when matter fields intervene. 

In this paper, we relied specifically on the axion-${\rm U}(1)$ gauge field model \eqref{AfM} for sourcing gravitational waves: this model involves, in addition to the usual scalar field driving inflation, a spectator sector including a gauge field with ${\rm U}(1)$ symmetry directly coupled to an axion. 
We considered two choices for the rolling axion potential (M1 and M2 in \eqref{pots}), both capable of giving localized gauge field amplification at large/intermediate CMB scales. In Section \ref{s1p2}, we provided bounds on the parameter space of the model, and more specifically on the effective coupling $\xi_{*}$ (and $\lambda$) between the axion and the gauge field, as implied by self-consistency of the theory, i.e. validity of the perturbative regime and negligible backreaction from the gauge field quanta. The theoretical  bounds are summarized in \eqref{sumc} and the resultant available parameter space is shown in Fig. \ref{fig:c}.

In Section \ref{sec:observation}, we completed the analysis of the model by deriving upper bounds on $\xi_{*}$ from state-of-the-art CMB spectra, namely from the latest \textit{Planck} and BICEP/Keck data. We adopt for this purpose the frequentist profile likelihood approach, fully exploiting in this context its immunity to prior choices, model parametrization and volume effects, which instead are known to affect Bayesian estimates. We summarize in Fig. \ref{fig:upper_xi_vs_k} the upper bounds on $\xi_{*}$ from \textit{Planck} and BICEP/Keck data, for typical choices of the model parameters $\delta$ and $k_{*}$ which control, respectively, the width and the position in wavenumber space of the bump feature sourced by gauge fields in scalar and tensor spectra.

\begin{figure}[t!]
\begin{center}
\includegraphics[scale=0.62]{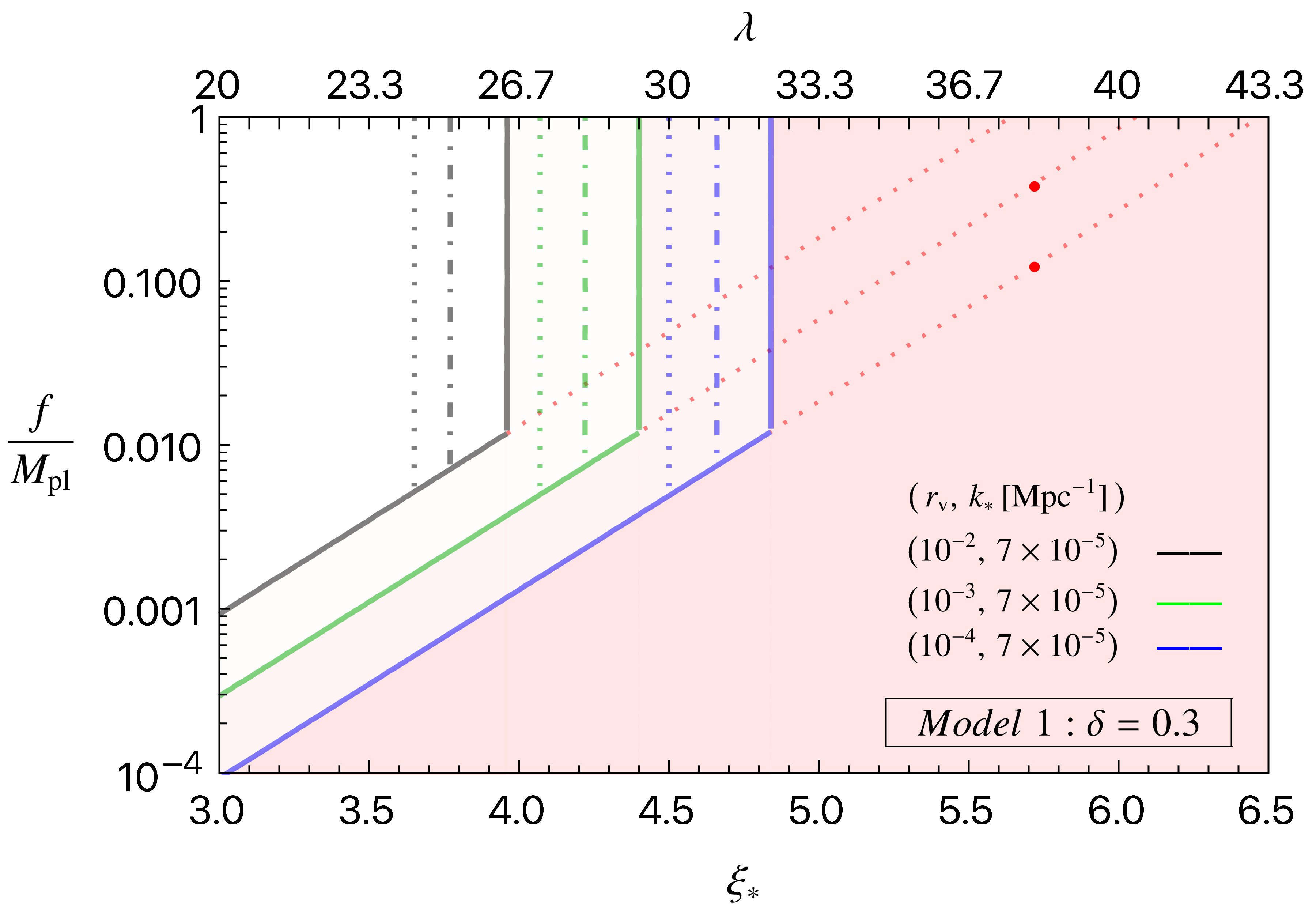}\includegraphics[scale=0.62]{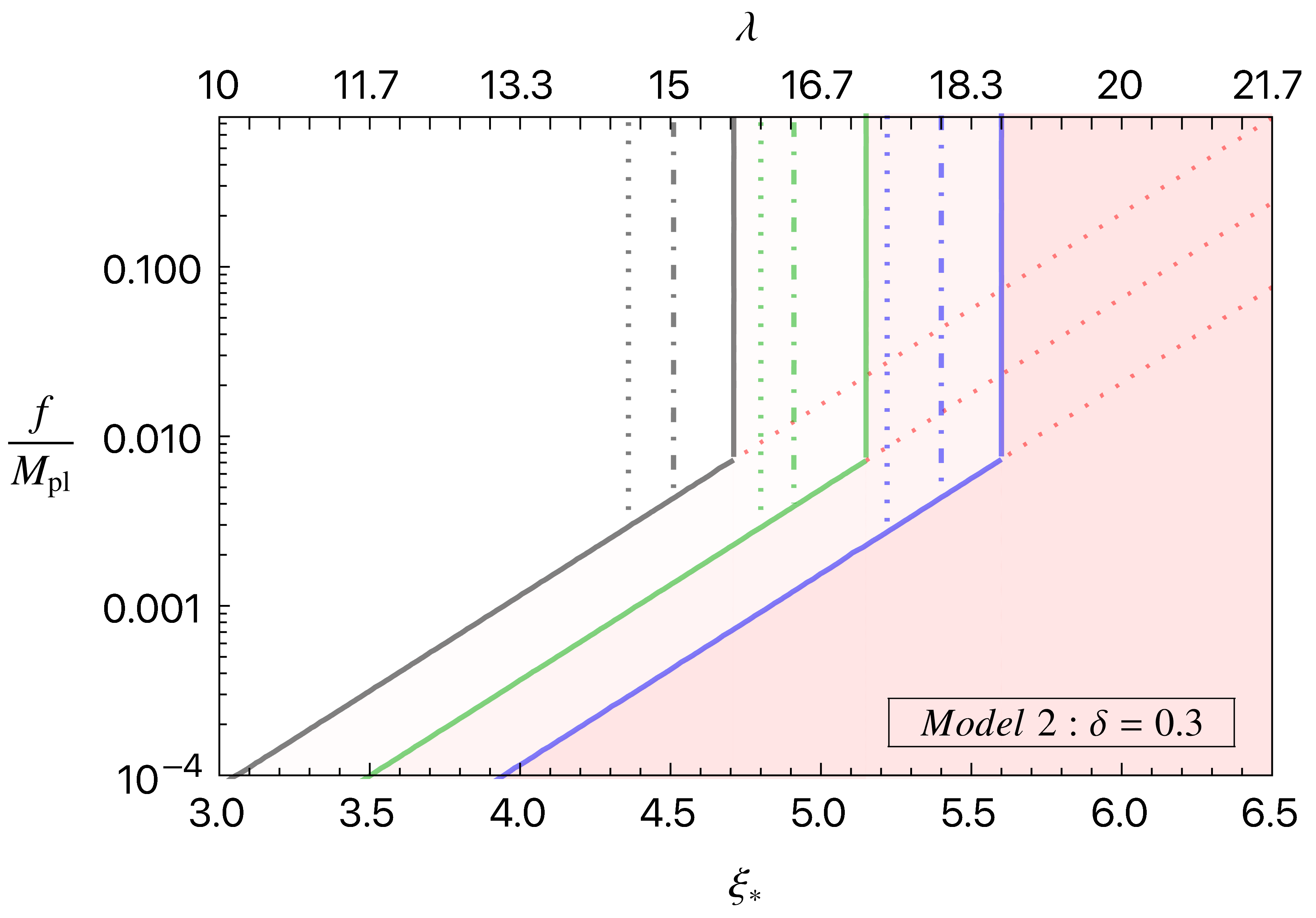}
\includegraphics[scale=0.62]{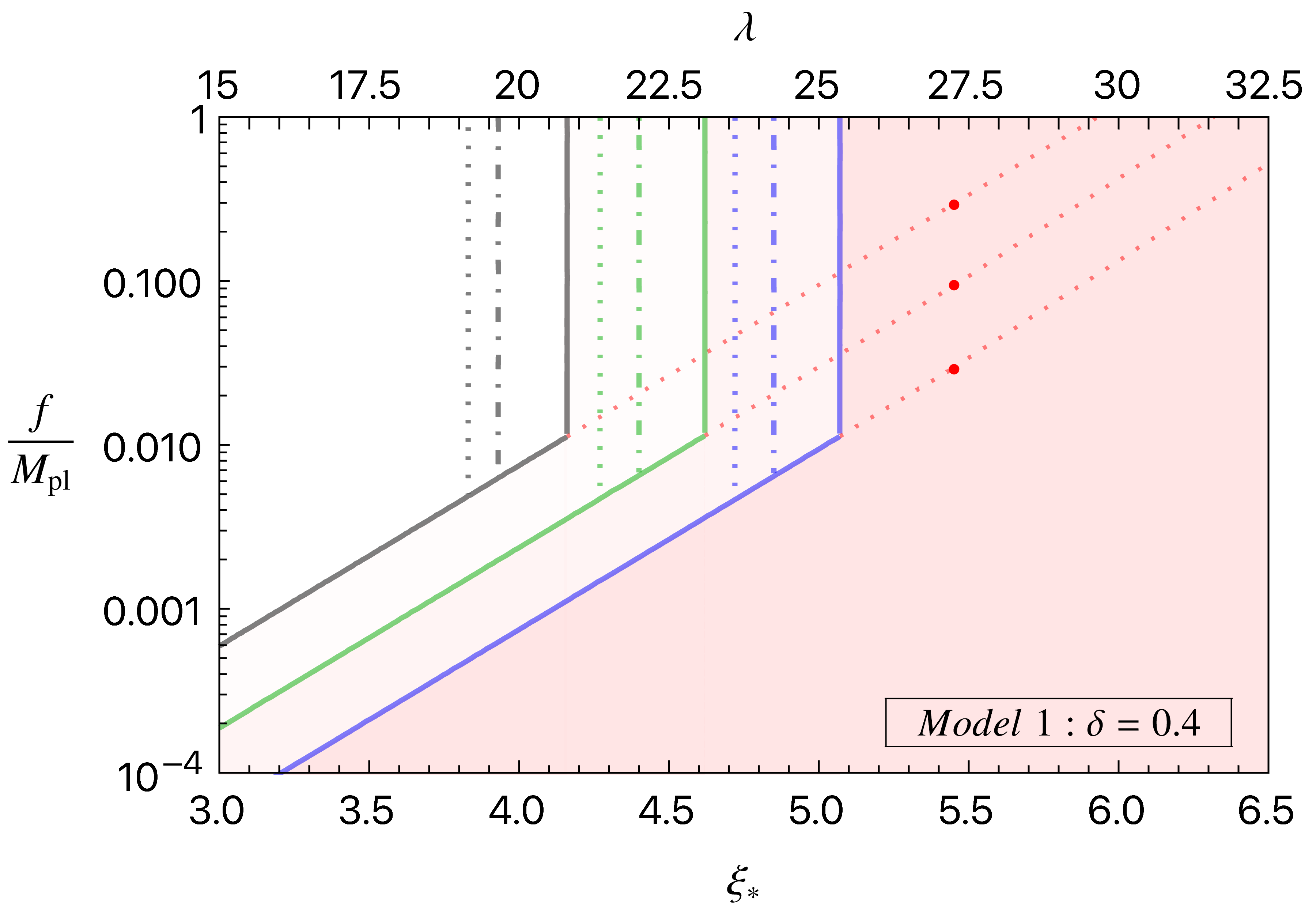}\includegraphics[scale=0.62]{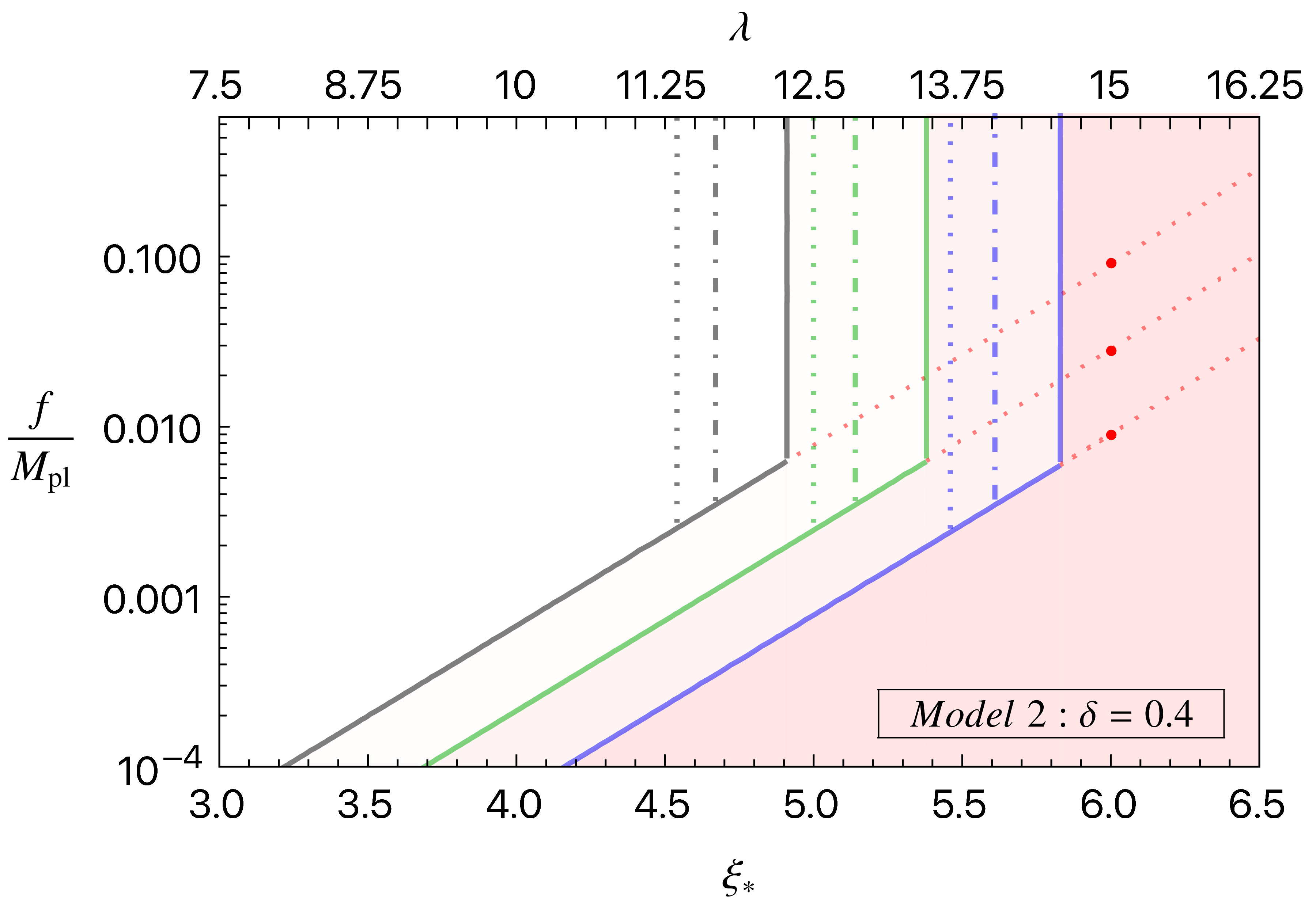}
\includegraphics[scale=0.62]{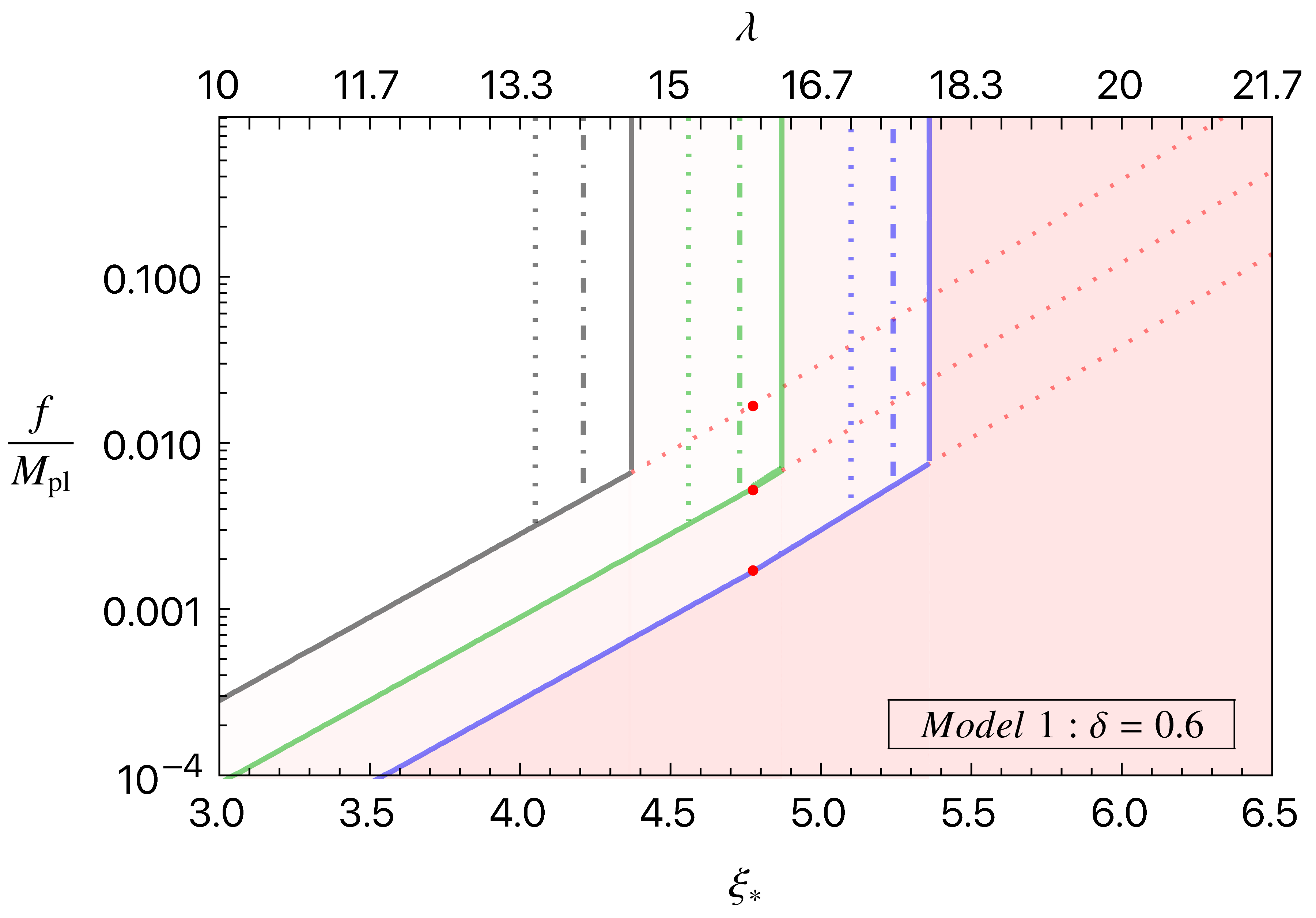}\includegraphics[scale=0.62]{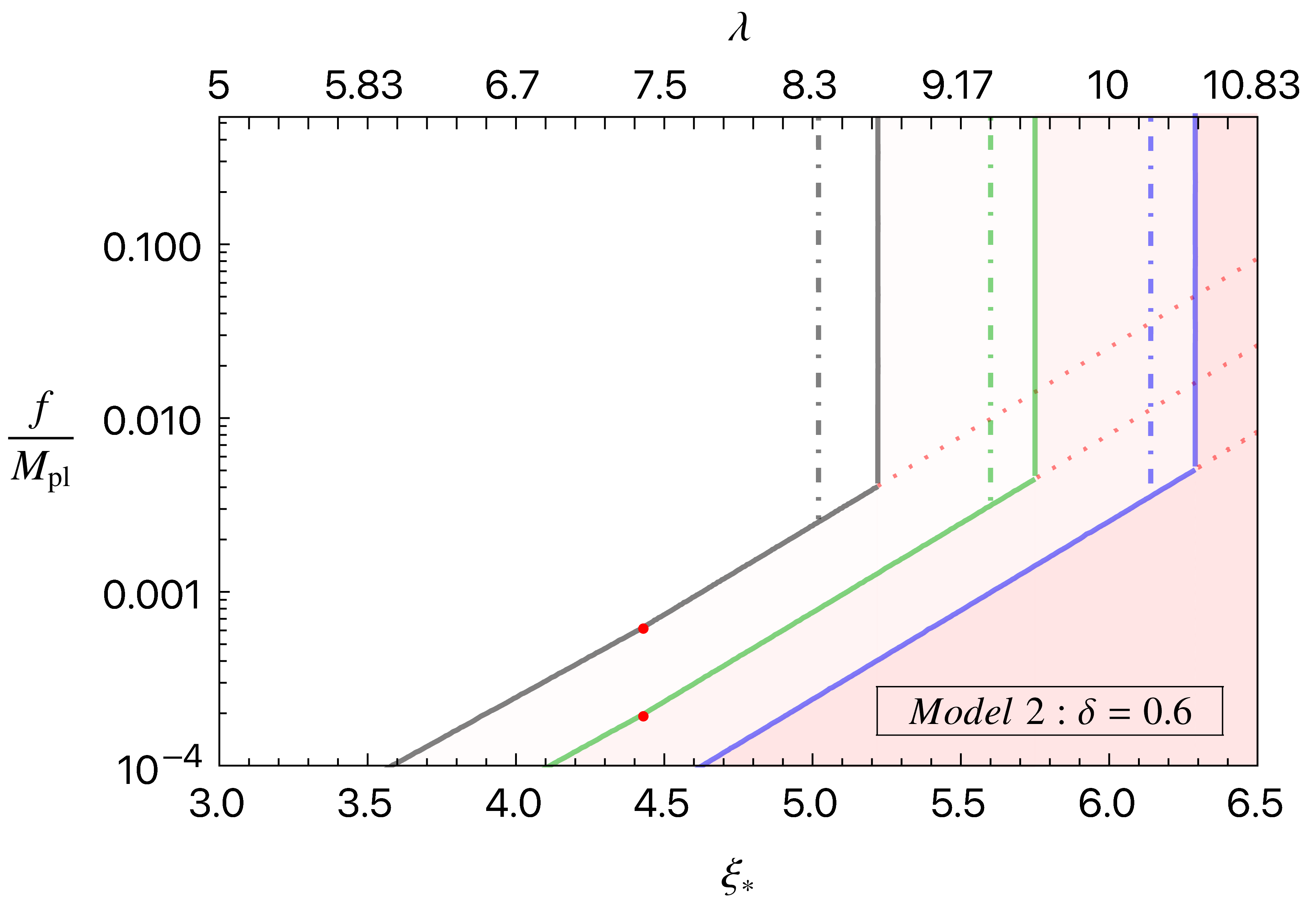}
\end{center}
\caption{Viable parameter space $f/\Mpl - \xi_* (\lambda)$ of the spectator axion-${\rm U}(1)$ gauge field model consistent with backreaction, perturbativity and observational data (see Sections \ref{s1p2} and \ref{3p3} and compare to Fig. \ref{fig:c}) for the M1 (Left) and M2 (Right) model with $\delta = 0.3,0.4,0.6$. 
For each choice of $r_{\rm v}$ shown (black, green and blue lines corresponding to $r_{\rm v} = 10^{-2},\,10^{-3},\, 10^{-4}$, respectively), the allowed parameter space, consistent with the limiting $\xi_*$ values from \textit{Planck} and BICEP/Keck data, is to the left of the solid, dot-dashed or dotted lines for $k_* = 7 \times 10^{-5}$, $5 \times 10^{-4}$ and $5 \times 10^{-3}\, {\rm Mpc}^{-1}$, respectively. \label{fig:finc}}
\end{figure}

\begin{figure}
    \centering
    \includegraphics[scale=0.5]{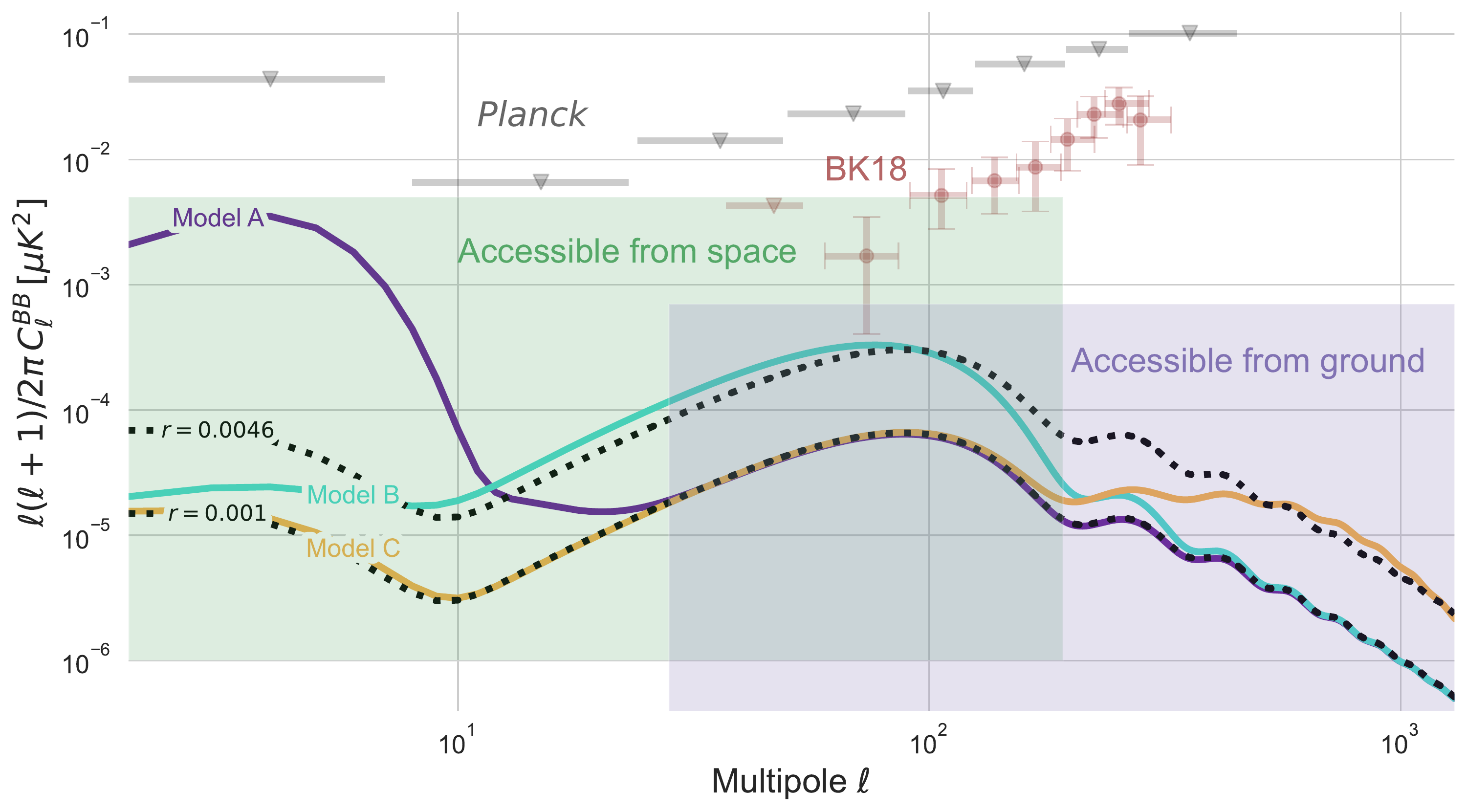}
    \caption{Benefits of measuring $B$-modes on large scales with a space mission (e.g. LiteBIRD) and on small scales with ground-based experiment (e.g. CMB-S4). The green and purple-shaded areas represent the range of multipoles to which each kind of experiment is typically sensitive to. Dotted black lines show the single-field slow-roll prediction for $r=0.0046$ (predicted by the Starobinsky model \cite{Starobinsky:1979ty, Starobinsky:1980te}) and for $r=0.001$ (close to detectability limit for future $B$-mode probes). Solid lines show theoretical CMB spectra candidates evaluated at $\xi_{*,\,limit}$. Specifically, \textit{Model A} (M2 model: $k_{*}=7\times 10^{-5}\,{\rm Mpc}^{-1}$, $r_{\rm v} = 0.001$ and $\delta=0.4$, solid purple) is indistinguishable at $\ell>30$ from the vacuum prediction for $r=0.001$, while featuring a very distinctive reionization bump. \textit{Model B} (M1 model: $k_{*}=5\times 10^{-4}\,{\rm Mpc}^{-1}$, $r_{\rm v} = 0.001$, $\delta=0.2$, light blue) features instead a recombination bump very similar to the one for $r=0.0046$, but with a quite different behavior from the vacuum prediction at reionization bump and smaller scales. Finally, \textit{Model C} (M1 model: $k_{*}=5\times 10^{-3}\,{\rm Mpc}^{-1}$, $r_{\rm v} = 0.001$, $\delta=0.6$, light orange) is indistinguishable from $r=0.001$ in the whole multipole range accessible from space, but has a distinctive bump feature at $\ell>200$, making a ground-based mission necessary to distinguish it from the standard slow-roll prediction. For reference, we also show the 95\% C.L. error bars and upper limits from \textit{Planck} (in gray) and BK18 (in red) data.
    }
    \label{fig:experimentalist}
\end{figure}

\smallskip
\noindent{\bf Reduced viable parameter space by \textit{Planck} and BK18.} 
The observational upper limits we obtained on $\xi_*$ can further tighten the parameter space of the model \eqref{AfM}, and are competitive with the theoretical bounds presented in Section \ref{s1p2} for the self-consistency of our approach. 
In particular, as discussed in Section \ref{3p3}, the \textit{Planck} and BICEP/Keck data limits the size of the effective coupling $\xi_*$ and hence the height of the maximally chiral, scale-dependent tensor perturbations sourced by the gauge fields at scales $k_*$ relevant for CMB observations. Including these observational bounds leads to a further reduction of the available parameter space consistent with backreaction and perturbativity bounds, in the $f/\Mpl-\xi_* (\lambda)$ plane for the spectator axion gauge field model. For $\delta = 0.3,0.4,0.6$, we superimpose these observational constraints with the theoretical bounds (see Fig. \ref{fig:c}) and show the resulting parameter space in Fig. \ref{fig:finc}. Comparing with the perturbativity + backreaction constraints presented in Fig. \ref{fig:c}, we can clearly observe that the available parameter space shrinks from a large triangle at fixed $r_{\rm v}$ to a smaller right trapezoid by the observational constraints on $\xi_*$ (i.e. $\xi_{*,limit}$), shown by the vertical lines corresponding to bounds obtained at $k_* = 7 \times 10^{-5}$ (solid), $k_* = 5 \times 10^{-4}$ (dot-dashed) and $k_* = 5 \times 10^{-3}$ (dotted)  ${\rm Mpc}^{-1}$. As can be confirmed from these plots and from our discussion in the previous section, tightest limits on the area of available parameter space of the models \eqref{AfM} come from the smallest scales\footnote{The only exception is the second Model (M2) with $\delta=0.6$, for which the physical peak $k_{\rm p}$ of the sourced signals occurs at scales where observational constraints by \textit{Planck} and BK18 are weak (see section \ref{3p3} and footnote 19). For this reason, in Figure \ref{fig:finc} we did not include the observational constraints on M2 with $\delta=0.6$ at $k_* = 5 \times 10^{-3}\, {\rm Mpc}^{-1}$ (see the bottom right panel).} $k_* = 5 \times 10^{-3}\, {\rm Mpc}^{-1}$  while the allowed region gradually enlarges towards larger scales at fixed $r_{\rm v}$. Similar to our discussion in Section \ref{s1p2}, for choices of $r_{\rm v}$ smaller than what is presented in Fig. \ref{fig:finc}, a larger parameter space can in principle be made available. It is worth stressing that such cases correspond to inflaton sectors endowed with flatter scalar potential $V(\phi)$ where $V'\sim \sqrt{\epsilon_\phi}\, V /\Mpl \sim \sqrt{r_{\rm v}}\, V/\Mpl $. Within the available parameter space presented in Figure \ref{fig:finc}, the total tensor-to-scalar ratio $r_*$ can be inferred from Figure \ref{fig:rstar}. In particular, we clearly observe that a sizeable sourced contribution to $r_*$ from gauge fields is viable while complying with CMB data at all $k_*$ values we consider.

\smallskip
\noindent{\bf The path ahead: relevance of a $B$-mode satellite mission.} As we discussed in Section \ref{3p3}, the current observational constraints on the model \eqref{AfM} are mainly driven by temperature spectra (i.e. from the sourced contribution to scalar fluctuations). $B$-mode polarization data at large/intermediate scales from \textit{Planck} and BICEP/Keck are indeed weakly constrained, and therefore have a minor effect on the model bounds (Figures \ref{fig:TT_vs_TEB} and \ref{fig:TT_vs_TEB_M2}). Large-scale temperature data are already cosmic variance-limited in the \textit{Planck} dataset, so sensitivity to (large/intermediate scale) polarization must be improved to better constrain the axion-${\rm U}(1)$ gauge field model. In particular, we argue that a $B$-mode satellite mission with access to the large and intermediate CMB scales (i.e. the ones pertaining to the reionization bump), such as LiteBIRD, would have unique benefits in distinguishing a vacuum-generated SGWB from a sourced one in the model under consideration \cite{Campeti:2019ylm, LiteBIRD:2022cnt}.
Ground-based experiments, such as BICEP/Keck considered in this paper or the planned high-sensitivity CMB-S4, indeed, cannot access the multipoles $\ell \lesssim 30$, as they typically have much smaller sky coverage, compared to the almost full-sky measurements available from space, and are affected by Earth's atmospheric contamination at the largest scales. 
Nonetheless, future ground-based experiments would still be highly beneficial and complementary to a satellite mission: high-sensitivity measurements of intermediate scales $B$-modes would help discriminating between vacuum and sourced origins of GWs for $k_{*} \gtrsim 5 \times 10^{-4}\, {\rm Mpc}^{-1}$. In Fig. \ref{fig:experimentalist}, we illustrate the potential benefits of measuring both large and small scale $B$-modes, with a space mission and a ground-based experiment, respectively.   
A full-sky space mission would also be necessary to obtain improved measurements of $EB$ and $TB$ parity-violating correlations at the largest scales, which are non-vanishing for gauge-sourced SGWB production during inflation, as discussed in Section \ref{sec:intro}. Furthermore, LiteBIRD will also greatly improve limits on tensor non-Gaussianity at large scales, making a signal of order $\mathcal{O}(1)$ potentially detectable in tensor-tensor-tensor equilateral (i.e. $f_{NL}^{ttt,{\rm eq}}$) and squeezed ($f_{NL}^{ttt,{\rm sq}}$) configurations \cite{Shiraishi:2019yux}.

\acknowledgments

We are greatly indebted to Eiichiro Komatsu for invaluable help and guidance during the work. PC’s work was supported by the Deutsche Forschungsgemeinschaft (DFG, German Research Foundation) under Germany's Excellence Strategy - EXC-2094 - 390783311. The work of O\"O is supported by the European Structural and Investment Funds and the Czech Ministry of Education, Youth and Sports (Project CoGraDS-CZ.02.1.01/0.0/0.0/15003/0000437).
{IO acknowledges the support from JSPS Overseas Research Fellowship.
This work is supported by the JSPS KAKENHI Grants No. JP19K14718 (MS) and JP20H05859 (IO and MS).} 
The numerical analyses in this work have been supported by the Max Planck Computing and Data Facility (MPCDF) computer clusters \textit{Cobra}, \textit{Freya} and \textit{Raven}. 
MS acknowledges the Center for Computational Astrophysics, National Astronomical Observatory of Japan, for providing the computing resources of Cray XC50.

\begin{appendix}
\section{Fitting functions for the sourced power spectra}
In this appendix, we provide fitting functions for the peak amplitude $f_{2, j}^{c}$, position $x_{2, j}^{c}$ and width $\sigma_{2, j}$ of the sourced scalar and tensor 2-pt signals in eq. \eqref{SC}. In particular, up to quadratic order, the dependence of these functions on the effective coupling $\xi_*$ is provided in Tables \ref{tab:fit1}-\ref{tab:fit4} for progressively faster rolling spectator axion corresponding to $\delta = \{0.2,0.3,0.4,0.6\}$.
\begin{table}[h!]
\begin{center}
\begin{tabular}{| c | c | c | c |}
\hline
$\{i,j\}$&$\ln(f^c_{i,j}) \simeq $&$x^c_{i,j} \simeq$&$\sigma_{i,j} \simeq$\\
\hline
$\{2,+\}$ & \scalebox{0.95}{$-13.89 +9.94\, \xi_* + 0.1082\, \xi_* ^2$ }&\scalebox{0.95}{$2.42 + 0.083\,\xi_* + 0.0278\, \xi_*^2$}&\scalebox{0.95}{$1.48 -0.187\,\xi_* + 0.0122\,\xi_* ^2 $}\\\hline
$\{2,-\}$ & \scalebox{0.95}{$-7.67 +9.92\, \xi_* + 0.1094\, \xi_* ^2$}&\scalebox{0.95}{$6.02 + 0.205\,\xi_* + 0.0578\, \xi_*^2$}&\scalebox{0.96}{$1.44 -0.204\,\xi_* + 0.0131\,\xi_* ^2 $}\\\hline
$\{2,\mathcal{R}\}$&\scalebox{0.95}{$ -5.30 + 9.95\,\xi_*+ 0.1059\,\xi_*^2$}&\scalebox{0.95}{$3.65 + 0.252\, \xi_* + 0.0306\, \xi_* ^2$}&\scalebox{0.95}{$1.18 -0.081\, \xi_* + 0.0021\, \xi_* ^2$}\\\hline
\hline
\hline
$\{2,+\}$ & \scalebox{0.95}{$-22.40 +10.26\, \xi_* +0.0757\, \xi_* ^2$} & \scalebox{0.95}{$6.25 - 0.783\,\xi_* + 0.0980\,\xi_* ^2$}& \scalebox{0.95}{$ 1.11 -0.122\,\xi_* + 0.0080\,\xi_* ^2$} \\\hline
$\{2,-\}$ &\scalebox{0.95}{$ -16.13 + 10.23\,\xi_*+ 0.0771\,\xi_*^2$} & \scalebox{0.95}{$15.63 -2.106\, \xi_* + 0.2425\, \xi_* ^2$}&\scalebox{0.95}{$1.05 -0.138\, \xi_* + 0.0087\, \xi_* ^2$}\\\hline
$\{2,\mathcal{R}\}$& \scalebox{0.95}{$ -15.67 + 10.31\,\xi_*+ 0.0745\,\xi_*^2$} & \scalebox{0.95}{$14.08 - 2.274\, \xi_* + 0.2277\, \xi_* ^2$} & \scalebox{0.95}{$1.09 -0.127\, \xi_* + 0.0080\, \xi_* ^2$} \\\hline
\end{tabular}
\caption{\label{tab:fit1} $\xi_*$ dependence of peak height, width and the location of the Gaussian template \eqref{SC} in the Models M1 (top three rows) and M2 (bottom three rows) for $\delta = 0.2$ and ${3\leq \xi_*\leq 6.5}$. }
\end{center}			
\end{table}
\begin{table}[h!]
\begin{center}
\begin{tabular}{| c | c | c | c |}
\hline
$\{i,j\}$&$\ln(f^c_{i,j}) \simeq $&$x^c_{i,j} \simeq $ &$\sigma_{i,j} \simeq $ \\
\hline
$\{2,+\}$ & \scalebox{0.95}{$-13.93 +9.80\, \xi_* + 0.0826\, \xi_* ^2$} & \scalebox{0.95}{$1.65 + 0.251\,\xi_* + 0.0188\, \xi_*^2$}&\scalebox{0.95}{$1.12 -0.127\,\xi_* + 0.0084\,\xi_* ^2 $}\\\hline
$\{2,-\}$ & \scalebox{0.95}{$-7.70 +9.77\, \xi_* + 0.0845\, \xi_* ^2$} & \scalebox{0.95}{$4.20 + 0.508\,\xi_* + 0.0428\, \xi_*^2$}&\scalebox{0.95}{$1.06 -0.149\,\xi_* + 0.0096\,\xi_* ^2 $}\\\hline
$\{2,\mathcal{R}\}$&\scalebox{0.95}{$ -6.13 + 9.76,\xi_*+ 0.0834\,\xi_*^2$} &\scalebox{0.95}{$2.57 + 0.408\, \xi_* + 0.0221\, \xi_* ^2$}&\scalebox{0.95}{$1.11 -0.137\, \xi_* + 0.0089\, \xi_* ^2$}\\\hline
\hline
\hline
$\{2,+\}$ & \scalebox{0.95}{$-21.09 +9.97\, \xi_* +0.0441\, \xi_* ^2$} & \scalebox{0.95}{$3.19 - 0.009\,\xi_* + 0.0412\,\xi_* ^2$ }&\scalebox{0.95}{$ 0.91 -0.091\,\xi_* + 0.0060\,\xi_* ^2$ }\\\hline
$\{2,-\}$ &\scalebox{0.95}{$ -14.85 + 9.94\,\xi_*+ 0.0461\,\xi_*^2$} & \scalebox{0.95}{$7.87 -0.201\, \xi_* + 0.1025\, \xi_* ^2$}&\scalebox{0.95}{$0.83 -0.111\, \xi_* + 0.0071\, \xi_* ^2$}\\\hline
$\{2,\mathcal{R}\}$&\scalebox{0.95}{$ -15.12 + 10.1\,\xi_*+ 0.0390\,\xi_*^2$} &\scalebox{0.95}{$6.71 -0.435\, \xi_* + 0.0887\, \xi_* ^2$}&\scalebox{0.95}{$0.90 -0.103\, \xi_* + 0.0068\, \xi_* ^2$} \\\hline
\end{tabular}
\caption{\label{tab:fit2} Same as previous Table except for the choice $\delta = 0.3$.} 
\end{center}			
\end{table}
\begin{table}[h!]
\begin{center}
\begin{tabular}{| c | c | c | c |}
\hline
$\{i,j\}$&$\ln(f^c_{i,j}) \simeq $&$x^c_{i,j} \simeq $ &$\sigma_{i,j} \simeq $ \\
\hline
$\{2,+\}$ & \scalebox{0.95}{$-12.87 +9.28\, \xi_* + 0.0844\, \xi_* ^2$} & \scalebox{0.95}{$1.38 + 0.339\,\xi_* + 0.0133\, \xi_*^2$}&\scalebox{0.95}{$0.95 -0.099\,\xi_* + 0.0065\,\xi_* ^2 $}\\\hline
$\{2,-\}$ & \scalebox{0.95}{$-7.15 +9.37\, \xi_* + 0.0774\, \xi_* ^2$} & \scalebox{0.95}{$3.05 + 0.818\,\xi_* + 0.0211\, \xi_*^2$}&\scalebox{0.95}{$0.88 -0.120\,\xi_* + 0.0077\,\xi_* ^2 $}\\\hline
$\{2,\mathcal{R}\}$&\scalebox{0.95}{$ -6.27 + 9.37,\xi_*+ 0.0753\,\xi_*^2$} &\scalebox{0.95}{$1.89 + 0.580\, \xi_* + 0.0096\, \xi_* ^2$}&\scalebox{0.95}{$0.92 -0.105\, \xi_* + 0.0068\, \xi_* ^2$}\\\hline
\hline
\hline
$\{2,+\}$ & \scalebox{0.95}{$-19.82 +9.46\, \xi_* +0.0319\, \xi_* ^2$} & \scalebox{0.95}{$2.32 + 0.214\,\xi_* + 0.0260\,\xi_* ^2$ }&\scalebox{0.95}{$ 0.82 -0.076\,\xi_* + 0.0052\,\xi_* ^2$ }\\\hline
$\{2,-\}$ &\scalebox{0.95}{$ -13.62 + 9.43\,\xi_*+ 0.0337\,\xi_*^2$} & \scalebox{0.95}{$5.63 +0.344\, \xi_* + 0.0660\, \xi_* ^2$}&\scalebox{0.95}{$0.72 -0.096\, \xi_* + 0.0061\, \xi_* ^2$}\\\hline
$\{2,\mathcal{R}\}$&\scalebox{0.95}{$ -14.54 + 9.68\,\xi_*+ 0.0195\,\xi_*^2$} &\scalebox{0.95}{$4.61 +0.070\, \xi_* + 0.0529\, \xi_* ^2$}&\scalebox{0.95}{$0.77 -0.079\, \xi_* + 0.0051\, \xi_* ^2$} \\\hline
\end{tabular}
\caption{\label{tab:fit3} Same as previous Table except for the choice $\delta = 0.4$. }
\end{center}			
\end{table}
\begin{table}[h!]
\begin{center}
\begin{tabular}{| c | c | c | c |}
\hline
$\{i,j\}$&$\ln(f^c_{i,j}) \simeq $&$x^c_{i,j} \simeq $ &$\sigma_{i,j} \simeq $ \\
\hline
$\{2,+\}$& \scalebox{0.95}{$-11.96 +8.58\, \xi_* + 0.0608\, \xi_* ^2$} &\scalebox{0.95}{$1.18 + 0.408\,\xi_* + 0.0098\, \xi_*^2$}&\scalebox{0.95}{$0.80 -0.072\,\xi_* + 0.0049\,\xi_* ^2 $}\\\hline
$\{2,-\}$& \scalebox{0.95}{$-6.36 +8.67\, \xi_* + 0.0542\, \xi_* ^2$} & \scalebox{0.95}{$4.62 + 2.224\,\xi_* + 0.0992\, \xi_*^2$}&\scalebox{0.95}{$0.86 -0.113\,\xi_* + 0.0072\,\xi_* ^2 $}\\\hline
$\{2,\mathcal{R}\}$&\scalebox{0.95}{$ -6.42 + 8.66,\xi_*+ 0.0527\,\xi_*^2$} &\scalebox{0.95}{$2.59 + 1.384\, \xi_* + 0.0498\, \xi_* ^2$}&\scalebox{0.95}{$0.90 -0.088\, \xi_* + 0.0058\, \xi_* ^2$} \\\hline
\hline
\hline
$\{2,+\}$ &\scalebox{0.95}{$-18.02 +8.49\, \xi_* +0.0154\, \xi_* ^2$}& \scalebox{0.95}{$1.81 + 0.346\,\xi_* + 0.0174\,\xi_* ^2$} &\scalebox{0.95}{ $ 0.74 -0.062\,\xi_* + 0.0043\,\xi_* ^2$} \\\hline
$\{2,-\}$ &\scalebox{0.95}{$ -11.89 + 8.46\,\xi_*+ 0.0170\,\xi_*^2$} & \scalebox{0.95}{$9.30 +1.536\, \xi_* + 0.1829\, \xi_* ^2$}&\scalebox{0.95}{$0.76 -0.099\, \xi_* + 0.0063\, \xi_* ^2$}\\\hline
$\{2,\mathcal{R}\}$&\scalebox{0.95}{$ -13.18 + 8.16\,\xi_*- 0.0103\,\xi_*^2$} &\scalebox{0.95}{$2.65 +0.557\, \xi_* + 0.0164\, \xi_* ^2$}&\scalebox{0.95}{$0.60 -0.041\, \xi_* + 0.0024\, \xi_* ^2$} \\\hline
\end{tabular}
\caption{\label{tab:fit4} Same as previous Table except for the choice $\delta = 0.6$.}
\end{center}			
\end{table}
\end{appendix}

\newpage
\bibliographystyle{JHEP}
\bibliography{biblio}

\end{document}